\documentclass[prl,twocolumn, superscriptaddress]{revtex4}
\usepackage{graphicx,amsmath,amssymb,amsxtra}
\usepackage{color}
\def\bd{\begin{displaymath}}
\def\be{\begin{equation}}
\def\ed{\end{displaymath}}
\def\ee{\end{equation}}
\def\bsub{\begin{subequations}}
\def\esub{\end{subequations}}

\newcommand{\Eq}[1]{Eq.~(\ref{#1})}
\newcommand{\Fig}[1]{Fig.~\ref{#1}}

\newcommand \idmat{ I } % identity matrix
\newcommand \nn{ \nonumber }

\makeatletter
\newcommand*\dashline{\rotatebox[origin=c]{90}{$\dabar@\dabar@\dabar@$}}
\makeatother

\begin{document}
\title{Entanglement Entropy and Topological Order in Resonating Valence-Bond Quantum Spin Liquids}

\author{Julia Wildeboer}

\affiliation{
Department of Physics and National High Magnetic Field Laboratory,
Florida State University, Tallahassee, Florida 32310, USA}
\affiliation{
Perimeter Institute for Theoretical Physics,
Waterloo, Ontario, N2L 2Y5, Canada}

\author{Alexander Seidel}

\affiliation{
Department of Physics and Center for Materials Innovation,
Washington University, St. Louis, Missouri 63130, USA}
\affiliation{Max-Planck-Institut f\"ur Physik komplexer Systeme, 
N\"othnitzer Str. 38, 01187 Dresden, Germany}

\author{Roger G. Melko}

\affiliation{
Perimeter Institute for Theoretical Physics,
Waterloo, Ontario, N2L 2Y5, Canada}
\affiliation{
Department of Physics and Astronomy,
University of Waterloo, Ontario, N2L 3G1, Canada}

\begin{abstract} 
On the triangular and kagome lattices, short-ranged resonating valence bond (RVB) wave functions  
can be sampled without the sign problem using a recently-developed Pfaffian Monte Carlo scheme. 
In this paper, we study the Renyi entanglement entropy in these wave functions using a replica-trick method. 
Using various spatial bipartitions, including the Levin-Wen construction, our finite-size scaled Renyi entropy 
gives a topological contribution consistent with $\gamma = \text{ln}(2)$, as expected for a gapped $\mathbb{Z}_{2}$ 
quantum spin liquid. 
We prove that the mutual statistics are consistent with the toric code anyon model 
and rule out any other quasiparticle statistics such as the double semion model. 
\end{abstract}
%***************************

\maketitle

%***************************
{\em Introduction.} -- 
Two-dimensional frustrated quantum antiferromagnets can harbor a phase of matter  
called a quantum spin liquid; a state with no conventional symmetry but emergent,  
topological order \cite{wen_paper,WenTopo}. 
These phases are unique in that they exhibit gapped fractionalized quasiparticle excitations 
with exotic quantum statistics and ground state degeneracies on topologically non-trivial surfaces \cite{LeonQSL}. 
Although there is strong incentive to identify minimal theoretical models which possess topologically ordered phases, 
the fact that strong correlations are a crucial ingredient means that numerical methods necessarily play a large role.  
Numerical studies suffer several serious challenges. First, the vast majority of Hamiltonians and wave functions that 
may harbor candidate quantum spin liquid states are also afflicted with the ``sign problem'', precluding 
study by large-scale quantum Monte Carlo (QMC) \cite{Designer}. 
Also, the absence of a local order parameter in a quantum spin liquid 
means that topological order must be characterized through more refined techniques, such as 
universal scaling terms in the entanglement entropy - the {\it topological} entanglement entropy (TEE) 
$-\gamma$ \cite{lw,KP}. 
Since $\gamma$ is sub-leading to the diverging ``area-law'', it can be challenging to extract 
in numerical simulations \cite{Isakov,Zhang2,Grover,sheng}. 
Finally, distinct topological phases defined by different emergent quasiparticles can have the same TEE. 
To distinguish, one must rely on the modular $\mathcal{U}$ and $\mathcal{S}$-matrices, which encode information on the quasiparticle 
statistics of the underlying topological phase \cite{zhang_ashvin,Cincio}. 
\begin{figure}%[t]
\centering
\includegraphics[width=9cm]{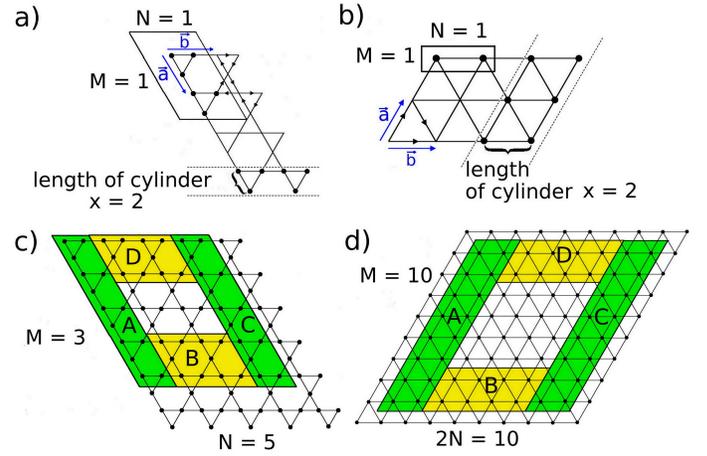}
\caption{ \label{unit}
a) The 6-site unit cell of the kagome lattice and the phase convention for singlets living on the lattice links. 
The length of the cylindrical regions that the entanglement entropy $S_{2}$ is measured over is shown as $x$. 
b) The 2-site unit cell, the respective phase convention and the definition of $x$ for the triangular lattice. 
c) and d) show the Levin-Wen areas $A,B,C$ and $D$ for the kagome lattice of size $(M,N) = (3,5)$ amounting to $90$ sites, and the 
$10 \times 10$-triangular lattice, respectively. 
The TEE $-\gamma$ is obtained by a superposition of the EE of four areas: 
$-2\gamma = S_{ABCD} - S_{ABC} - S_{ADC} + S_{AC}$. 
 }
\end{figure}

In this Letter, we analyze the Renyi entanglement entropy of the short-ranged 
spin-$\frac{1}{2}$ 
resonating valence bond (RVB) 
wave function on the kagome and the triangular lattice. 
Recently Ref.~\cite{wildeboer12} introduced a sign-problem free Pfaffian Monte Carlo scheme that can be used 
to produce unbiased samples of the singlet wave function, 
making it possible to evaluate local operators and their correlation functions. 
That work demonstrated that the RVB wave function on these two frustrated lattices has no local order parameter, 
and is gapped, consistent with expectations for an $SU(2)$-invariant quantum spin liquid. 
Here, we use the Pfaffian Monte Carlo technique to calculate the TEE, which explicitly shows $\gamma = \text{ln}(2)$, as expected 
for a $\mathbb{Z}_2$ topologically-ordered phase. 
Further, we prove that the mutual statistics are consistent with the toric code anyon model in both the triangular and kagome RVB states, 
ruling out any other underlying anyon models such as the double semion. 
\begin{figure}%[t]
\centering
\includegraphics[width=8.75cm]{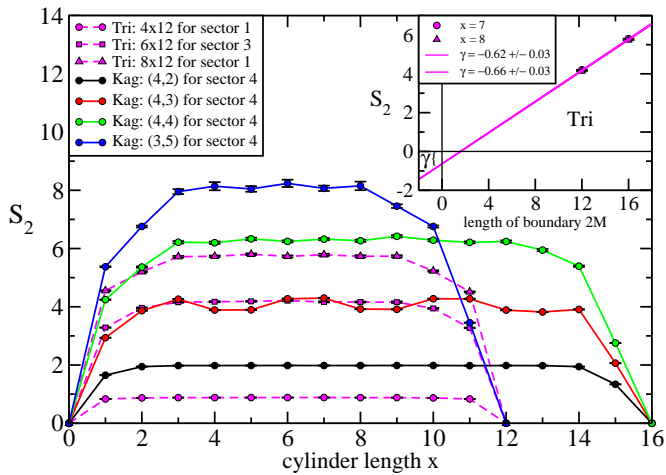}
\caption{ \label{M3456810_N12}
The Renyi entropy $S_{2}$ for triangular (Tri) lattices of size $M \times 2N = 4,6,8 \times 12$ and kagome (Kag) lattices 
for various sizes $(M,N)$ 
with total number of lattice sites $6MN$. 
There are four possible topological sectors for each wave function on a torus, each of which can be isolated independently in the Pfaffian MC. 
For the triangular lattice, the inset shows linear fits using $S_{2}$ from $M \times 2N = 6,8\times 12$ for lengths $x = 7,8$ 
in order to extract $\gamma^{\prime}$ for a single sector. 
In total, we average over values stemming from fits for $4 \leq x \leq 8$, and obtain an average value of $-\gamma^{\prime} = -0.64 \pm 0.07$. 
 }
\end{figure}
 
{\em RVB Wave Functions, Entanglement, and QMC.} -- 
The RVB wave functions were conceived by Anderson 40 years ago \cite{anderson} 
for their variational appeal in demonstrating spin liquid physics. 
The simplest, nearest-neighbor RVB state represents a stable phase only on non-bipartite 
lattices. 
Although not typically discussed as ground states of explicit local Hamiltonians, RVB wave functions 
do sometimes allow for the construction of a local parent Hamiltonian, as notable in 
particular on the kagome lattice \cite{seidel09}. 
The uniqueness of the RVB-ground states, modulo a topological degeneracy, 
(demonstrated in Refs.~\cite{schuch2, ZWS12}) 
establishes that the Hamiltonian in Ref.~\cite{seidel09} truly stabilizes the RVB state. 
In this work, we will directly consider the RVB wave function, defined via 
\begin{eqnarray}\label{eq4}
|RVB\rangle = \sum_D |D\rangle \,. 
\end{eqnarray}
Here, $D$ goes over all possible pairings of a given lattice into nearest neighbor pairs (``dimer coverings''). 
Each site of the lattice is equipped with a spin-$\frac{1}{2}$ degree of freedom. 
For each dimer covering $D$, $|D\rangle$ denotes a state where each pair of lattice sites of the covering 
forms a singlet, where a sign convention is used that %usually 
corresponds to an orientation of nearest neighbor links (see \Fig{unit}a \& b). 
We note that the wave function \Eq{eq4} has a fourfold topological degeneracy on the torus for the 
kagome and triangular lattices. 

Like in any quantum wave function, the properties of an RVB state can be investigated through its 
bipartite entanglement entropy, where the lattice is divided into a region $A$ and its complement $B$. 
The Renyi entropy of order $n$ is defined as 
$S_{n} = \text{ln Tr}(\rho_{A}^{n})/(1-n)$, 
where $\rho_{A} = \text{Tr}_{B}|\Psi \rangle \langle \Psi|$ is the reduced density matrix of region $A$. 
Ground states of local Hamiltonians are known to exhibit a {\it area law} scaling in region size, 
which in two dimensions can generically be written as, 
$S_n(\rho_{A}) = \alpha_n L_{A} - \gamma + \cdots $ \cite{Eisert}. 
Here, the leading term is dependent on the ``area'' (or boundary) of region $A$. 
The second term, the topological entanglement entropy (TEE) $-\gamma$ \cite{lw,KP}, 
is characterized by the total quantum dimension $\mathcal{D}$, which is defined through the 
quantum dimensions of the individual quasiparticles $d_{i}$ of the underlying theory: $\mathcal{D} = \sqrt{\sum_{i} d_{i}^2}$ \cite{lw,KP,fradkin}. 
Conventionally ordered phases have $\mathcal{D} = 1$, 
while topologically ordered phases have $\mathcal{D} > 1$ with the 
TEE given by $-\gamma = -\text{ln}(\mathcal{D})$. 

Note that in the case where the area $A$ has at least one non-contractible boundary, such as a cylinder (see \Fig{unit}a \& b), $\gamma$ becomes 
state-dependent. 
As shown in Ref.~\cite{zhang_ashvin}, if one expresses 
any state in the basis of the {\it minimum entropy states} (MES-states), 
$| \Psi_{\alpha} \rangle = \sum_{j} c_{j} | \Xi_{j} \rangle$, 
then the sub-leading constant to the area law from a two-cylinder cut is, 
\begin{eqnarray}\label{gamma_torus2} 
\gamma^{\prime}(\{p_{j}\}) = 2\gamma + \text{ln}\bigg(\sum_{j} \frac{p_{j}^{2}}{d_{j}^{2}} \bigg) \;. 
\end{eqnarray}
for $S_2$, where $p_{j} = |c_{j}|^2$. We further discuss MES-states in the results to follow. 

In contrast to bipartite lattices \cite{sutherland,anderson_docout,VBQMC,alet,tang,ju,stephan,punk2015quantum}, 
RVB states on non-bipartite lattices are not amenable to valence-bond QMC; they have been studied previously by PEPs representations \cite{PEPS1,PEPS2,PEPS3}, 
but should also be accessible to QMC if a sign-problem free sampling method can be constructed \cite{becca,YangYao,wildeboer12}. 
Here, we investigate entanglement properties of the spin-$\frac{1}{2}$ RVB 
wave functions using the variational Pfaffian MC scheme for lattices of up to $128$ sites. 
Note that the Pfaffian MC scheme allows one to project onto each topological sector, and every linear combination thereof. 
We will use this feature in the following results. 
To obtain the second Renyi entropy $S_{2}$ for contractible and noncontractible regions, 
we employ the standard QMC replica-trick \cite{hastings,RatioT}. We refer to Refs.~\cite{SM,wildeboer12} for more details on the method. 

{\em Measurements of TEE.} -- 
We begin by calculating the TEE using boundaries for region $A$ that are contractible around the toroidal lattice. 
To isolate $\gamma$, we perform a Levin-Wen bipartition \cite{lw}, 
which was successfully used previously to detect a $\mathbb{Z}_2$ quantum spin liquid 
using QMC simulations on toroidal lattices of restricted finite-size \cite{Isakov}. 
We obtain data for such bipartitions on both a triangular RVB of size $M \times 2N = 10 \times 10$ and two kagome RVBs 
with $(M,N) = (3,5),(3,6)$ amounting to $90$ and $108$ sites, respectively. 
The triangular lattice and the $(3,5)$-kagome geometries are shown in \Fig{unit}, 
which also shows the Levin-Wen regions $A,B,C,D$ used to obtain $\gamma$ \cite{lw}. 
For the $(3,6)$-kagome, the $A(C)$ regions are the same as in $(3,5)$, whereas the regions $B(D)$ are one link longer in $N$-direction than in 
$(3,5)$. 
Using this procedure, the triangular lattice gives 
$-\gamma \approx -0.80 \pm 0.2$, while for the kagome, we end up with $-\gamma = -0.89 \pm 0.22, -0.74 \pm 0.34$ for $(3,5)$ and $(3,6)$, respectively. 

To improve accuracy, we now consider regions with non-contractible boundaries. 
We examine a triangular lattice RVB for fixed $2N = 12$ and $M = 4, 6, 8$, 
and a kagome lattice RVB of size $6MN$ with $M = 4$ and $N = 2, 3, 4$. 
We subsequently calculate the Renyi entropy $S_{2}$ for cylindrical bipartitions (see \Fig{unit}). 
As the cylinder length increases, $S_{2}$ quickly saturates (\Fig{M3456810_N12}). 
This type of behavior is consistent with the system having a gap. 
We point out that, as expected, several curves in \Fig{M3456810_N12} exhibit finite size effects, 
manifest clearly in the different $S_{2}$ for different topological sectors. 
This can be seen in \Fig{kagome_MES_nonMES}, 7 and 8 \cite{SM}. 

Figures \ref{kagome_MES_nonMES} and \ref{kagome_sec14_supp} demonstrate that, for large enough system sizes, $S_{2}$ does not depend on the 
topological sector for these two wave functions. 
The four topological sectors $|\Psi_{i} \rangle$ of each wave function can be distinguished by their two quantum numbers, 
$i = \{ee, eo, oe, oo\}$, 
which are the even $(e)$ or odd $(o)$ number of dimers %(transition-graph loops) 
cut along the two directions $\vec{a}$ and $\vec{b}$. 
One can make use of special linear combinations of these topological sectors 
to devise another method of determining the TEE $-\gamma$. 
Here, we choose as a compatible ansatz for $\mathbb{Z}_{2}$ spin liquids the minimal-entangled states (MES) obtained for the 
toric code 
and the dimer model on kagome/triangular lattices 
\cite{zhang_ashvin,SM} and examine its properties. 
These MES-states for cuts along $\vec{b}$ are, up to a phase $e^{i\Phi_{j}}$, 
\begin{eqnarray}\label{MES_Z2} 
&&| \Xi_{1,2}\rangle = \frac{1}{\sqrt{2}} (| \Psi_{ee} \rangle \pm | \Psi_{eo} \rangle)\;,  \nonumber \\
&&| \Xi_{3,4}\rangle = \frac{1}{\sqrt{2}} (| \Psi_{oe} \rangle \pm | \Psi_{oo} \rangle)\;.
\end{eqnarray} 
We apply this ansatz to our Pfaffian QMC data. 
 First, consider slices (of constant cylinder length $x$) 
through the triangular lattice to obtain a plot of 
$S_{2}(L_{2M}) = \alpha L_{2M} - \gamma^{\prime}$, as in the inset of \Fig{M3456810_N12}. 
The intercept of this plot is $-\gamma^{\prime}$ which is now the state dependent TEE; 
we numerically extract $-\gamma^{\prime} = -0.64 \pm 0.07$. 
We can thus obtain the TEE using Eq.~\eqref{gamma_torus2}. First note that, as 
seen in Fig.~\ref{kagome_sec14_supp} of the SM,  
the entropy $S_2$ does not depend on which MES-state is used (within error bars), which 
implies that  $d_j$ is also the same for all four MES-states.  Since every theory, Abelian and non-Abelian, 
contains (at least) one quasiparticle with quantum dimension unity, we conclude 
that all quasiparticle dimensions are necessarily $d_j = 1$.  Since the $p_j$ are fixed to be 
$1/4$ by our ansatz Eq.~\eqref{MES_Z2}, then for the single sector  
plotted in \Fig{M3456810_N12},  
$\gamma^{\prime} = 2\gamma + \text{ln}({1}/{4} + {1}/{4})$.  Thus, we conclude that 
$\gamma = 0.67 \pm 0.04$ 
consistent with $\mathbb{Z}_2$ topological order. 

Next we turn to the kagome lattice RVB. As seen in \Fig{kagome_MES_nonMES} (inset) and Fig. $8$ in the SM \cite{SM}, 
it takes a system of size $(M,N) = (4,3)$ to 
eliminate finite-size effects and reach agreement of $S_{2}$ of all four topological sectors within error bars. 
Since larger system sizes can become computationally expensive, 
an extraction of $\gamma^{\prime}$ using the procedure of \Fig{M3456810_N12} becomes more difficult. 
Alternatively, as discussed in Ref.~\cite{zhang_ashvin}, 
linear superpositions of different MES states can be used to extract  
all information about the topological 
order from our measurement. Specifically, the linear combination of all four MES-states, 
\begin{align}\label{wrong_sec}
 & |\Sigma_{1,2}\rangle = \frac{1}{\sqrt{2}}(| \Psi_{ee} \rangle + | \Psi_{oe(oo)} \rangle) \nonumber \\
& = \frac{1}{2} \bigg(| \Xi_{1}\rangle + | \Xi_{2}\rangle + | \Xi_{3}\rangle 
\pm | \Xi_{4}\rangle\bigg),
\end{align}
will have $\gamma^{\prime} = 0$ according to Eq.~\eqref{gamma_torus2}. 
Here, index $1$ ($2$) corresponds to $oe$ ($oo$), and $+$ ($-$) in the second line. 

\begin{figure}[t]
\centering
\includegraphics[width=8.75cm]{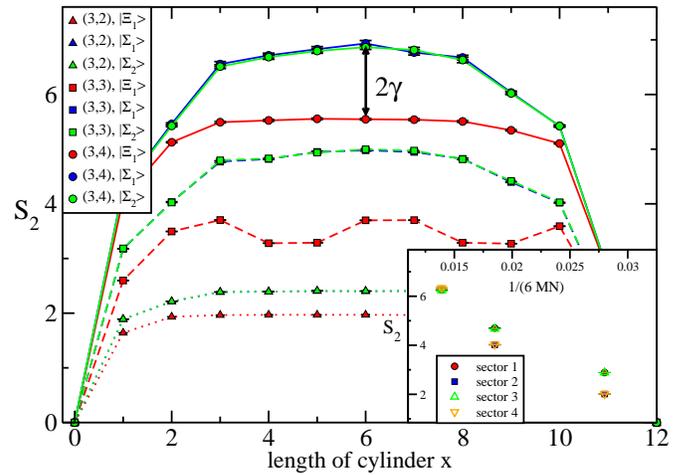}
\caption{ \label{kagome_MES_nonMES}
$S_{2}$ for kagome-lattice MES-state $(|\Xi_{1}\rangle)$ which is a specific superposition of two 
topological sectors for different lattice sizes and for two respective nonMES-states $(|\Sigma_{1,2}\rangle)$. 
The difference between $(M,N) = (3,4)$ for the two nonMES-states compared to the MES-state has to be $2\gamma = 2\text{ln}(2) \approx 1.38$. 
If we average over differences obtained for lengths $x = 5, 6, 7$, we get $-\gamma^{\prime} = -2\gamma = -1.36 \pm 0.13$. 
The inset shows a comparison of $S_2$ for cylindrical regions between sectors $1, 2, 3, 4$ on the kagome lattice. 
We observe that for large enough system sizes $(M,N) = (3,4)$ amounting to $72$ lattice sites, 
the EE $S_{2}$ is the same for all four sectors within errors. 
 }
\end{figure}
We investigate the behavior of $S_{2}$ for the MES-state, Eq.~\eqref{MES_Z2}, and the nonMES-states, Eq.~\eqref{wrong_sec}, 
for the kagome lattice RVB. 
We first note that the numerical data of \Fig{kagome_MES_nonMES} (Figs. $7,8$ in the SM) suggests $S_{2}$ 
to be independent of MES-state (once finite-size effects are accounted for). 
Since $S_{2}$ is similarly the same across all four topological sectors, this implies that $\gamma^{\prime}$ is the same for all $| \Xi_{i}\rangle$ 
and for all $|\Psi_{i} \rangle$, respectively. 
The latter means that each of the four quasi-particles belonging to the four MES-states have the same quantum dimension. 
Since every phase, Abelian and non-Abelian, has (at least) one quasi-particle with quantum dimension $d = 1$, this implies that all 
quasi-particles have $d_{i} = 1$, indicating the Abelian nature of the phase. 
We calculate the difference in $S_{2}$ between MES \eqref{MES_Z2} and nonMES \eqref{wrong_sec} states by 
performing an average over cylinder lengths $x = 4, \ldots, 8$, and obtain $-\gamma^{\prime} = -1.36 \pm 0.13$. 
This matches the expectation from Eq.~\eqref{gamma_torus2} that this difference should be $-2\gamma \approx -1.38$, 
confirming that the MES ansatz is also correct for the kagome-lattice RVB. 

{\em Quasiparticle Statistics.} -- 
We point out that the numerical confirmation of the MES-states ansatz \Eq{MES_Z2} 
essentially determines the topological order of our system, and in particular 
distinguishes between toric code and double semion topological order as we now explain. 
We consider 
the matrices $\mathcal{S}$ and $\mathcal{U}$, which 
describe the quasiparticle statistics of the system and correspond 
to modular transformations of the same name at the level of the effective 
field theory. 
In a microscopic lattice model, the corresponding transformations cannot 
necessarily be realized as 
discrete symmetry operations. 
However, for both the kagome and the triangular lattice, 
the transformation corresponding to $\mathcal{U}\mathcal{S}^{-1}$ 
is realized as the symmetry under a $\pi/3$-rotation \cite{SM}, 
as long as the lattice dimensions are chosen to be of the form 
$(M,N) = (M,2M)$ or $(M, 2N) = (2N,2N)$ for the kagome or triangular, respectively. 
Up to a phase ambiguity \cite{zhang_ashvin}, 
the matrix elements $V_{ij}=\langle\Xi_{i}|R_{\pi/3}|\Xi_{j}\rangle$ 
are thus equal to those of the matrix $\mathcal{US}^{-1}$, where $R_{\pi/3}$ 
represents the $\pi/3$-rotation. We therefore must have 
$V_{ij}=D^{\dagger} \mathcal{US}^{-1} D$, where $D$ is a diagonal matrix of phases $D_{jj} = e^{i\Phi_{j}}$ corresponding to the 
phase ambiguity. 
$V_{ij}$ is easily calculated from \Eq{MES_Z2} by working out the transformation properties 
of the states $|\Psi_{\alpha,\beta}\rangle$ under rotation \cite{poilblanc2011competing,SM}. 
It is manifestly real, as is $\mathcal{US}^{-1}$ for the toric code, 
and we find agreement for $D_{jj} = 1$ for all $jj$'s. 
In contrast, for the double semion model, $\mathcal{US}^{-1}$, 
while having the same eigenvalues as in the toric code case, we note that $\mathcal{U}$ 
has some purely imaginary diagonal entries. Therefore, in the double semion case, 
$D^{\dagger} \mathcal{US}^{-1} D$ must have imaginary entries for any 
choice of $D$, and agreement with our MES-states cannot be achieved. 

Thus, the MES-states we identified demonstrate the underlying 
quasiparticle statistics to be consistent with the toric code model, ruling out any other statistics, 
in particular 
double semion statistics. 

{\em Conclusion.} -- 
In this work, we have used a sign-problem free Pfaffian quantum Monte Carlo (QMC) 
to calculate the second Renyi entropy of the nearest-neighbor RVB wave function 
on the triangular and kagome lattices. 
Through a bipartition of each lattice into 
Levin-Wen \cite{lw} regions, and cylindrical regions, 
we confirm that the topological entanglement entropy (TEE) 
is consistent with $-\gamma = -\text{ln}(2)$, the 
value for a $\mathbb{Z}_{2}$ quantum spin liquid. 
Finite-size scaling of the two-cylinder Renyi entropy for the triangular lattice and comparisons between $S_{2}$ for different wave functions 
in MES-basis for the kagome, 
confirm the ansatz MES-states taken for a $\mathbb{Z}_{2}$ topological gauge structure. 
Further, we identify the nature of the anyonic quasiparticles to be of toric code type, 
by explicitly showing that our numerically confirmed MES-states ansatz leads to the modular $\mathcal{US}^{-1}$-matrix of the 
toric code statistics 
and rules out any other quasiparticle statistics including double semion statistics. 

This work serves as an important example that all aspects of quantum spin liquid behavior, 
from the initial demonstration of the liquid nature \cite{wildeboer12} to the characterization of 
the emergent gauge structure through the TEE, to the full determination of the 
underlying statistics and braiding of fractional quasiparticle excitations, 
can be performed with un-biased QMC techniques. 
Thus, the $SU(2)$-invariant RVB states on triangular and kagome lattices 
add to the growing list of wave functions and 
Hamiltonians that have been demonstrated to exist, and can be simulated in practice, 
on non-bipartite lattices without being vexed by the sign-problem \cite{Designer,Kaul2015}. 

Finally, we emphasize that our results rely crucially on the numerical extraction of the 
second Renyi entropy of the quantum ground state. For RVB wave functions (and all other many-body 
systems), the replica-trick method used here \cite{hastings} is the same as that 
employed in recent experiments on interacting $^{87}$Rb atoms in a one-dimensional optical lattice \cite{Islam}. 
Hence, the concepts and techniques used in this paper will be important for efforts to characterize 
topological order in synthetic quantum matter in the near future.

 \begin{acknowledgments}
 The authors are indebted to J.~Carrasquilla, Ch.~Herdman, and E.~M.~Stoudenmire for enlightening discussions. We are especially 
 indebted to L.~Cincio for several critical readings of the manuscript. 
 AS would like to thank K. Shtengel for insightful discussions. 
 Our MC codes are partially based upon the ALPS libraries \cite{alps1,alps2}. 
 This work has been supported by the National Science Foundation under NSF Grant No. DMR-1206781 (AS), 
 NSERC, the Canada Research Chair program, and the Perimeter Institute (PI) for Theoretical Physics. 
 Research at Perimeter Institute is supported by the Government of Canada through Industry Canada and 
 by the Province of Ontario through the Ministry of Research and Innovation. 
 JW is supported by the National High Magnetic Field Laboratory under NSF Cooperative Agreement No. DMR-0654118 and the State of Florida. \\[0.35cm]
 \end{acknowledgments}

%\newpage
%%%%%%%%%%%%%%%%%%%%%%%%%%%%%%%%%%%%%%%%%%%%%%
%\title{Supplemental material: Entanglement Entropy and Topological Order in Resonating Valence-Bond Quantum Spin Liquids}
\textbf{Supplemental material: Entanglement Entropy and Topological Order in Resonating Valence-Bond Quantum Spin Liquids}

\author{Julia Wildeboer}

\affiliation{
Department of Physics and National High Magnetic Field Laboratory,
Florida State University, Tallahassee, Florida 32310, USA}
\affiliation{
Perimeter Institute for Theoretical Physics, 
Waterloo, Ontario, N2L 2Y5, Canada}

\author{Alexander Seidel}

\affiliation{
Department of Physics and Center for Materials Innovation, 
Washington University, St. Louis, Missouri 63130, USA}
\affiliation{Max-Planck-Institut f\"ur Physik komplexer Systeme, 
N\"othnitzer Str. 38, 01187 Dresden, Germany}

\author{Roger G. Melko}

\affiliation{
Perimeter Institute for Theoretical Physics,
Waterloo, Ontario, N2L 2Y5, Canada}
\affiliation{
Department of Physics and Astronomy, 
University of Waterloo, Ontario, N2L 3G1, Canada}
%***************************

\maketitle
%***************************
\medskip 

%{ \bf Supplemental material: }\\
{\em 1) The Kasteleyn method for the triangular lattice.} --  
The close-packed hard-core dimer model can be solved on any planar lattice by using Pfaffian techniques \cite{kas}. 
The key ingredient for the Pfaffian technique is to place arrows on the links of the planar graph/lattice so that 
each plaquette is ``clockwise odd'', that is to say that the product of the orientations of the arrows around 
any even-length elementary plaquette traversed clockwise is odd. 
Subsequently, an antisymmetric matrix $A$ is formed. 

Kasteleyn’s theorem then states that for any planar graph, $A$ can be found, and that the partition function 
which is the number of dimer coverings \cite{NNote1} 
is given by the Pfaffian of the matrix $A$: 
\begin{eqnarray}
Z = \pm {\rm Pf}(A)\;.
 \label{open_pfaf}
\end{eqnarray}
However, \eqref{open_pfaf} is only valid for a system with open boundary conditions (OBC). 
If we work on a toroidal system, we will see that we need a total of four Pfaffians. 

\begin{figure}[t]
\centering
\includegraphics[width=9cm]{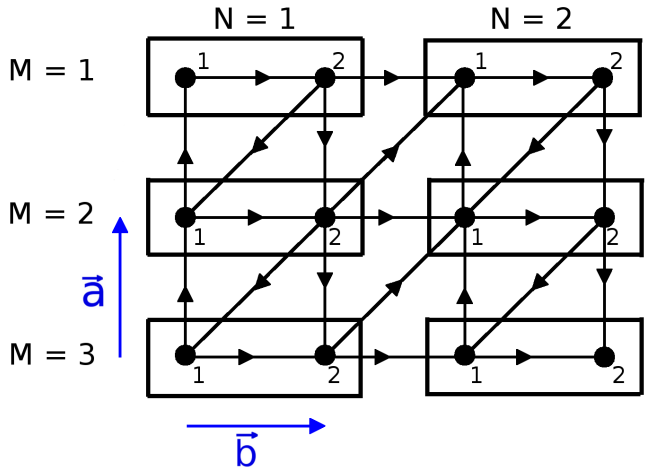}
\caption{ \label{figure0}
Shown is a triangular lattice with $M \times 2N = 3 \times 4 = 12$ sites with a Kasteleyn edge orientation. 
The unit cell consists of two lattice sites labeled 1 and 2, respectively. 
}
\end{figure}

Thus, we start by defining the four matrices $A_{i}$, $i = 1,2,3,4$ in case of the toroidal triangular lattice. 
The adopted Kasteleyn edge orientation and the unit cell of the triangular lattice containing $2$ sites, numbered $1$ and $2$, respectively, 
are shown in \Fig{figure0}. 
In the following, we consider a lattice of $M\times N$ unit cells having a total of $M \times 2N$ lattice sites, 
the case of $M = 3, 2N = 4$ is shown in \Fig{figure0}. 

The Pfaffian method \cite{kas} concerns with the evaluation of 
the $2MN\times 2MN$ antisymmetric Kasteleyn matrices $A$ written down according to 
edge weights and orientations (under specific boundary conditions) which can be 
read off from \Fig{figure0}, and by adopting the prescription 
\begin{eqnarray}
 A_{ij}=\left\{\begin{array}{lr}
 +w_{ij} & {\rm orientation\  from\ }i{\rm \ to\ }j\\
 -w_{ij} & {\rm orientation\  from\ }j{\rm \ to\ }i
 \end{array}\right.
 \label{wei}
\end{eqnarray}
where $w_{ij}$ is the weight of edge $ij$. In the following, we take all edge weights $w_{ij} = 1$. 

Kasteleyn\cite{kas} has shown that under periodic boundary conditions (PBC) the partition function 
$Z_{torus}$ 
is a linear combination of four Pfaffians Pf$(A_{i}), i = 1, \ldots, 4$, 
\begin{eqnarray}
 Z_{torus}=\frac{1}{2}\Big[-{\rm Pf}(A_1)+{\rm Pf}(A_2) 
  +{\rm Pf}(A_3)+{\rm Pf}(A_4)\Big]. \label{PBCGen1}
\end{eqnarray}
The four Pfaffians are specified by the Kasteleyn orientation 
of lattice edges with, or without, the reversal of arrows on edges 
connecting two opposite boundaries. 
A perusal of \Fig{figure0} and the use of the prescription \eqref{wei} 
lead to the four $2MN\times 2MN$ Kasteleyn matrices, 
\begin{eqnarray}
\begin{split}
%\begin{align}
  A_{1}  = & a_{0,0}\otimes \idmat_M\otimes \idmat_N
    + a_{0,1}    \otimes \idmat_M\otimes T_N
    - a_{0,1}^{T}\otimes \idmat_M\otimes T_N^{T} \\
 & + a_{1,0}    \otimes T_M     \otimes \idmat_N
    - a_{1,0}^{T}\otimes T_M^{T}     \otimes \idmat_N \\
 & + a_{1,1}    \otimes T_M     \otimes T_N^{T}
    - a_{1,1}^{T}\otimes T_M^{T} \otimes T_N \\
  A_{2}  = & a_{0,0}\otimes \idmat_M\otimes \idmat_N
    + a_{0,1}    \otimes \idmat_M\otimes H_N
    - a_{0,1}^{T}\otimes \idmat_M\otimes H_N^{T} \\
 & + a_{1,0}    \otimes T_M     \otimes \idmat_N
    - a_{1,0}^{T}\otimes T_M^{T}     \otimes \idmat_N \\
 & + a_{1,1}    \otimes T_M     \otimes H_N^{T}
    - a_{1,1}^{T}\otimes T_M^{T} \otimes H_N \\
  A_{3}  = & a_{0,0}\otimes \idmat_M\otimes \idmat_N
    + a_{0,1}    \otimes \idmat_M\otimes T_N
    - a_{0,1}^{T}\otimes \idmat_M\otimes T_N^{T} \\
 & + a_{1,0}    \otimes H_M     \otimes \idmat_N
    - a_{1,0}^{T}\otimes H_M^{T}     \otimes \idmat_N \\
 & + a_{1,1}    \otimes H_M     \otimes T_N^{T}
    - a_{1,1}^{T}\otimes H_M^{T} \otimes T_N \\
  A_{4}  = & a_{0,0}\otimes \idmat_M\otimes \idmat_N
    + a_{0,1}    \otimes \idmat_M\otimes H_N
    - a_{0,1}^{T}\otimes \idmat_M\otimes H_N^{T} \\
 & + a_{1,0}    \otimes H_M     \otimes \idmat_N
    - a_{1,0}^{T}\otimes H_M^{T}     \otimes \idmat_N \\
 & + a_{1,1}    \otimes H_M     \otimes H_N^{T}
    - a_{1,1}^{T}\otimes H_M^{T} \otimes H_N.
\end{split}
\label{fourmatrices}
\end{eqnarray}
%\end{align}
The superscripts $T$ denote transpose, $\otimes$ is the direct product, $\idmat_M$ is the $M\times M$ identity matrix, 
and $H_N$, $T_N$ are the $N\times N$ matrices 
%\begin{eqnarray}
%\begin{split}
\begin{align}
H_N=\begin{pmatrix}
 0     & 1     & 0     &\cdots & 0 \\
 0     & 0     & 1     &\cdots & 0 \\
\vdots &\vdots &\vdots &\ddots &\vdots \\
 0     & 0     & 0     &\cdots & 1 \\
-1     & 0     & 0     &\cdots & 0
\end{pmatrix} , \quad
T_N=\begin{pmatrix}
 0     & 1     & 0     &\cdots & 0 \\
 0     & 0     & 1     &\cdots & 0 \\
\vdots &\vdots &\vdots &\ddots &\vdots \\
 0     & 0     & 0     &\cdots & 1 \\
 1     & 0     & 0     &\cdots & 0
\end{pmatrix}.
\end{align}
% \end{split}
%\end{eqnarray}
and
\begin{eqnarray}
 a_{0,0} & = & \begin{pmatrix}
  \;\;\;0 &  1 \cr
 -1 &  0 \cr
\end{pmatrix},\quad
a_{0,1} = \begin{pmatrix}
 0 &  0 \cr
 1 &  0 \cr
\end{pmatrix}, \nn\\
a_{1,0} & = & \begin{pmatrix}
 -1 &  0 \cr
  \;\;\;1 &  1 \cr
\end{pmatrix} ,\quad
a_{1,1}   =  \begin{pmatrix}
 0 & -1 \cr
 0 &  \;\;\;0 \cr
\end{pmatrix}, \label{equ:amatrices}\\  \nn  \\
a_{-1,0} &=& - a_{1,0}^{T} ,\quad a_{0,-1}   =   -
a_{0,1}^{T},\quad a_{-1,-1}   =   - a_{1,1}^{T}. \nn
\end{eqnarray}
The need for four matrices is explained in the following. 

\begin{table}
    \begin{tabular}{  c | c | c | c | c }
    class of configurations & \multicolumn{4}{c}{sign of terms in ${\rm Pf}(A_{i})$ } \\
\cline{2-5}
                            & $\;\;A_1\;\;$ & $\;\;A_2\;\;$ & $\;\;A_3\;\;$ & $\;\;A_4\;\;$ \\ \hline \hline
    sector $1$: $(e,e)$        & +     & +     & +     & +     \\ \hline
    sector $2$: $(o,e)$        & -     & -     & +     & +     \\ \hline
    sector $3$: $(e,o)$        & -     & +     & -     & +     \\ \hline
    sector $4$: $(o,o)$        & -     & +     & +     & -     \\ \hline
    \end{tabular}
\caption{The counting rules for the dimer coverings belonging to the four sectors $(ee)$, $(eo)$, $(oe)$, $(oo)$.}
\label{tab:table2}
\end{table}
The single dimer coverings fall into four sectors that are distinguished 
by the even ($e$) or odd ($o$) number of dimer cuts of a loop along the two directions $\vec{a}$ and $\vec{b}$ of the torus. 
Each of the four Pfaffians only counts one or three types of the four sectors $(ee)$, $(eo)$, $(oe)$, $(oo)$, 
correctly (see Table ~\ref{tab:table2}). 
%\begin{table}
%    \begin{tabular}{  c | c | c | c | c }
%    class of configurations & \multicolumn{4}{c}{sign of terms in ${\rm Pf}(A_{i})$ } \\
%\cline{2-5}
%                            & $\;\;A_1\;\;$ & $\;\;A_2\;\;$ & $\;\;A_3\;\;$ & $\;\;A_4\;\;$ \\ \hline \hline
%    sector $1$: $(e,e)$        & +     & +     & +     & +     \\ \hline
%    sector $2$: $(o,e)$        & -     & -     & +     & +     \\ \hline
%    sector $3$: $(e,o)$        & -     & +     & -     & +     \\ \hline
%    sector $4$: $(o,o)$        & -     & +     & +     & -     \\ \hline
%    \end{tabular}
%\caption{The counting rules for the dimer coverings belonging to the four sectors $(ee)$, $(eo)$, $(oe)$, $(oo)$.} 
%\label{tab:table2}
%\end{table}
However, if we form the superposition \eqref{PBCGen1}, then all dimer coverings from all four sectors are 
counted with the correct sign. 
It is now also possible to project onto each sector: 
\begin{eqnarray}
Z_{1} &=& \frac{1}{4}({\rm Pf}(A_1) - {\rm Pf}(A_2) - {\rm Pf}(A_3) - {\rm Pf}(A_4)) \nonumber \\
Z_{2} &=& \frac{1}{4}(-{\rm Pf}(A_1) - {\rm Pf}(A_2) + {\rm Pf}(A_3) + {\rm Pf}(A_4)) \nonumber \\
Z_{3} &=& \frac{1}{4}(-{\rm Pf}(A_1) + {\rm Pf}(A_2) - {\rm Pf}(A_3) + {\rm Pf}(A_4))  \nonumber \\
Z_{4} &=& \frac{1}{4}(-{\rm Pf}(A_1) + {\rm Pf}(A_2) + {\rm Pf}(A_3) - {\rm Pf}(A_4))\;. \nonumber \\
&&
\end{eqnarray}
The partition sum for open boundary conditions can be obtained by taking any of the matrices $A_{i}$, $i = 1, \ldots, 4$, 
and setting the links in the $M(N)\times M(N)$ matrices that connect sites from opposite boundaries equal to zero. 

\begin{table}
    \begin{tabular}{ | c | c | c || r | r | r | r | r |}
    \hline
    $M$ & $N$ & $2MN$ & $Z_{torus}$  & $Z_{1}$   & $Z_{2}$   & $Z_{3}$   & $Z_{4}$   \\ \hline \hline
    3 &  2    & 12    & 344          & 92        & 80        & 92        & 80      \\ \hline
    4 &  2    & 16    & 1920         & 576       & 448       & 448       & 448     \\ \hline
    5 &  2    & 20    & 10608        & 2872      & 2432      & 2872      & 2432    \\ \hline
    6 &  2    & 24    & 59040        & 16720     & 12592     & 15696     & 14032   \\ \hline
    3 &  3    & 18    & 4480         & 1120      & 1120      & 1120      & 1120    \\ \hline
    4 &  3    & 24    & 59040        & 16720     & 15696     & 12592     & 14032   \\ \hline
    5 &  3    & 30    & 767776       & 191824    & 192064    & 191824    & 192064  \\ \hline
    3 &  4    & 24    & 58592        & 14576     & 14720     & 14576     & 14720   \\ \hline
    4 &  4    & 32    & 1826944      & 520256    & 512064    & 389184    & 405440  \\ \hline
    3 &  5    & 30    & 766528       & 191200    & 192064    & 191200    & 192064  \\ \hline
    3 &  6    & 36    & 10028288     & 2505344   & 2508800   & 2505344   & 2508800 \\ \hline
    \end{tabular}
\caption{Number of dimer coverings, $Z_{torus}$, of triangular lattices of with $2MN$ sites, 
with periodic boundary conditions. $Z_{i}$ gives the number of coverings for each topological sector $i$ with $i = 1, \ldots, 4$. 
Only those sizes ($M \geq 3$ and $N \geq 2$) are given in which any pair of sites is linked by at most one bond.}
\label{tab:table1}
\end{table}
To close this section, we remark that the analogous Pfaffian construction for the kagome lattice is available 
in Ref.\cite{wangwu} by Wu and Wang. Here, the unit cell consists of a total of $6$ lattice sites, the total number of sites 
is then $6MN$ (see Fig. $1$ in Letter). 
An analytic calculation reveals the full partition sum to be $Z_{torus} = 2 \times 4^{MN}$. 

Eventually, we give Tables ~\ref{tab:table1} and ~\ref{tab:table3} that list the full partition sum $Z_{torus}$ and the 
sector-wise partition sums $Z_{i}$, ($i = 1, \ldots, 4$), 
for small lattice sizes for the triangular and the kagome lattice. 
For the kagome lattice, the single sectors have the feature that they exactly have the same number of dimer coverings, 
$Z_{i} = \frac{1}{2} \times 4^{MN}$, for all pairs $(M,N)$ for $i = 1, \ldots, 4$. 

For the triangular lattice, we observe that in general different lattice sizes lead to different number of dimer coverings in the 
single sectors. 
However, it is possible to have $\frac{1}{4} Z_{torus}$ coverings in each sector for certain pairs $(M,N)$, one example being 
$(M,N) = (3,3)$ with $\frac{1}{4} Z_{torus} = 1120$ dimer coverings per sector. 
\begin{table}
    \begin{tabular}{ | c | c | c || r | r |}
    \hline
    $M$ & $N$ & $6MN$ & $Z_{torus}$ & $Z_{i} (i = 1, \ldots, 4)$ \\ \hline \hline
      1 &  2  & 12    & 32          & 8            \\ \hline
      2 &  2  & 24    & 512         & 128          \\ \hline
      2 &  3  & 36    & 8192        & 2048         \\ \hline
      2 &  4  & 48    & 131072      & 32768        \\ \hline
      2 &  5  & 60    & 2097152     & 524288       \\ \hline
      3 &  2  & 36    & 8192        & 2048         \\ \hline
      3 &  3  & 54    & 524888      & 131072       \\ \hline
      3 &  4  & 72    & 33554432    & 8388608      \\ \hline
      3 &  5  & 90    & 2147483648  & 536870912    \\ \hline
      3 &  6  & 108    & 137438953472    & 34359738368      \\ \hline
%      4 &  1  & 24    & 512             & 128             \\ \hline
      4 &  2  & 48    & 131072          & 32768           \\ \hline
      4 &  3  & 72    & 33554432        & 8388608         \\ \hline
      4 &  4  & 96    & 8589934592      & 2147483648      \\ \hline
      4 &  5  & 120   & 2199023255552   & 549755813888    \\ \hline
%      4 &  6  & 144   & 562949953421312 & 140737488355328 \\ \hline
%      4 &  8  & 192   & 36893488147419103232  & 9223372036854775808 \\ \hline
    \end{tabular}
\caption{Number of dimer coverings, $Z_{torus}$, of kagome lattices of with $6MN$ sites, 
with periodic boundary conditions. $Z_{i}$ gives the number of coverings for each topological sector $i$ with $i = 1, \ldots, 4$.} 
\label{tab:table3}
\end{table}

{\em 2) Entanglement and SWAP-operator for the RVB wave function.} -- 
We start by briefly reviewing the variational Pfaffian Monte Carlo scheme introduced in Ref.\cite{wildeboer12}. 
The scheme was introduced to overcome a sign problem that prevents the application of the valence bond Monte Carlo (VBMC) technique to the 
short-ranged RVB wave function on the non-bipartite kagome and triangular lattices. 
Only for bipartite lattices it is possible to endow each link of the lattice with a sign/phase convention that renders all overlaps between 
different singlet coverings positive. This has to be so, since the overlaps serve as weights in the VBMC. 
For non-bipartite lattices, such as the kagome and the triangular lattice, there exists no such sign convention. 

The idea is now the re-express the RVB wave function in a orthogonal basis, which does not suffer from a sign problem, 
rather than the non-orthogonal valence bond basis of \Eq{eq4}. 

First, we re-cast the RVB wave function \eqref{eq4} in terms of the Ising basis of local $S^{z}$ eigenstates. 
For a system consisting of an even number $N$ of spins/sites, we thus re-express \Eq{eq4} as: 
\begin{eqnarray}\label{wf1_supp}
| \text{RVB} \rangle = \sum_{I} a_{I} |I \rangle,
\end{eqnarray} 
where $|I \rangle = |m_{1}^{z}\rangle \otimes |m_{2}^{z}\rangle \otimes \ldots \otimes |m_{N}^{z}\rangle$ runs over the Ising basis. 
The newly appearing non-trivial amplitude $a_{I}$ can easily be calculated as the Pfaffian of an $N \times N$ matrix. 
This Pfaffian can be re-expressed as a $N/2 \times N/2$-determinant which can be evaluated in polynomial time. 

We point out that our state $|\Psi\rangle$ given as in \eqref{wf1_supp} 
is actually the sum over all four topological sectors: 
$|\Psi \rangle = \sum_{i=1}^{4} |\Psi_{i}\rangle$. 
We stress that we are able to restrict the sum in the state $|\Psi \rangle$ to any sector through a calculation of four 
specific Pfaffians (see Section $1)$). Subsequently, we are able to project onto any 
topological sector and onto any linear combinations of sectors when sampling the RVB wave function. 
More details on this issue are given in \cite{kas,wildeboer12} and in Section $1)$. 

Next, we adapt the standard QMC replica trick, the so-called SWAP-trick, and apply it to \eqref{wf1_supp} in order to calculate 
the Renyi entropy. 

The Renyi entropy of order $n$ is defined as 
\begin{equation}\label{eq3_supp} 
S_{n} = \frac{1}{1-n} \text{ln}\,\text{Tr}(\rho_{A}^{n}) \;. 
\end{equation}
In the limit $n \rightarrow 1$, the von Neumann entropy is recovered. 
The TEE $-\gamma$ does not depend on $n$. 

Since it is computationally expensive to calculate the von Neumann entropy, 
we proceed by calculating the second Renyi entropy. 
Here, the well-known QMC replica trick allows us to avoid the costly calculation of the reduced density matrix $\rho_{A}$ 
which is needed for the von Neumann entropy 
and to alternatively efficiently calculate the second Renyi entropy by obtaining the expectation 
value of the so-called $\text{SWAP}$-operator \cite{hastings,RatioT}.  
\begin{figure}[t]
\centering
\includegraphics[width=8.75cm]{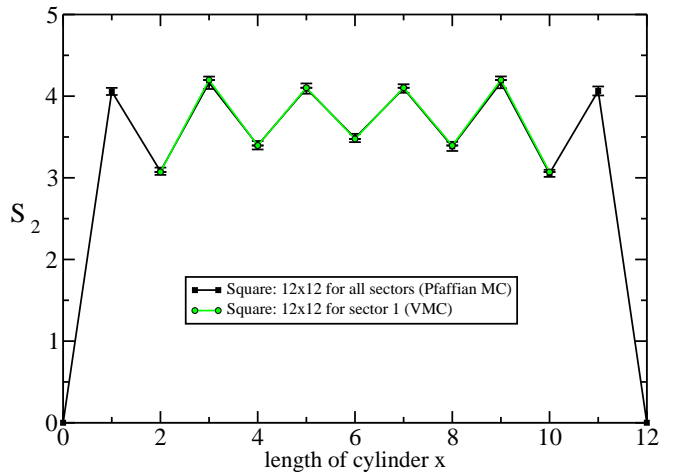}
\caption{ \label{compare_square_supp}
SQUARE: The EE $S_{2}$ over a cylindrical area $A$ is shown for the $12 \times 12$-square lattice. 
We do not observe any differences within error bars between $S_{2}$ for a specific sector and when calculated with the sum over all sectors. 
The ``all-sector''-data was obtained by Pfaffian Monte Carlo, while the ``single-sector''-data stems from a VBMC simulation. 
 }
\end{figure}
Using the SWAP-operator, calculating the second Renyi entropy 
over an area $A$ 
requires us to make two independent copies of the system, the total state of the doubled 
system is then $|\Psi \otimes \Psi^{\prime} \rangle = \sum_{I I^{\prime}} a_{I} a_{I^{\prime}} |I \otimes I^{\prime} \rangle$  
The $\text{SWAP}$-operator now exchanges the degrees of freedom in area $A$ between the two copies while leaving the degrees of freedom in 
the remaining area $B$ untouched: 
\begin{eqnarray}\label{sw_supp}
\text{SWAP} |I_{A} I_{B} \otimes I_{A}^{\prime} I_{B}^{\prime} \rangle = |I_{A}^{\prime} I_{B} \otimes I_{A} I_{B}^{\prime} \rangle \;. 
\end{eqnarray}

It can be shown that $\langle \text{SWAP} \rangle = \text{Tr}(\rho_{A}^{2})$ \cite{hastings} 
and consequently 
\begin{eqnarray}\label{s2_supp}
S_{2} = -\text{ln}(\langle \text{SWAP} \rangle)\;.
\end{eqnarray}
Using the RVB wave function \eqref{wf1_supp} 
for calculating the expectation value $\langle \text{SWAP} \rangle$, 
we arrive at the following: 
\begin{eqnarray}\label{eq11_supp}
\langle \text{SWAP} \rangle = \frac{\sum_{I I^{\prime}} |a_{I}|^{2} |a_{I^{\prime}}|^{2} 
\;\times\;
R}
{\sum_{I I^{\prime}} |a_{I}|^{2} |a_{I^{\prime}}|^{2} }
\end{eqnarray}
with the measurement/estimator 
\begin{eqnarray}\label{eq11a_supp}
R = \frac{a_{I^{s}} a_{I^{\prime,s}}}{a_{I} a_{I^{\prime}}}\;, 
\end{eqnarray}
where $I^{s}$ and $I^{\prime,s}$ refer to the Ising configurations of system and its copy {\em after} the SWAP-operation. 

It is clear from \eqref{eq11_supp} and \eqref{eq11a_supp} that the measurement depends on the state 
of the two systems after the exchange of the degree of freedoms. 

Since the number of Ising configurations grows exponentially with the size of the subsystem $A$, 
we have exponentially large fluctuations in the estimator $R$ and as it is implied by the area law, 
the convergence of the entanglement entropy becomes exponentially slow. 
Only an exponentially small part of the original Ising configuration will lead to a non-zero measurement of $R$. 
Most measurements will be zero. 
%\begin{figure}[t]
%\centering
%\includegraphics[width=8.75cm]{SQUARE_Roger_Julia_12x12_allsec-September7-1}
%\caption{ \label{compare_square_supp}
%SQUARE: The EE $S_{2}$ over a cylindrical area $A$ is shown for the $12 \times 12$-square lattice. 
%We do not observe any differences within error bars between $S_{2}$ for a specific sector and when calculated with the sum over all sectors. 
%The ``all-sector''-data was obtained by Pfaffian Monte Carlo, while the ``single-sector''-data stems from a VBMC simulation. 
% }
%\end{figure}
Thus, in order to combat the exponentially growing variance of the simple estimator $R$ in \eqref{eq11a_supp}, we employ a more refined 
re-weighting scheme from Ref.\cite{pei}. 
The new re-weighting scheme splits the expectation value $\langle \text{SWAP} \rangle$ into two parts 
\begin{eqnarray}\label{eq11b_supp} 
\langle \text{SWAP} \rangle = \langle \text{SWAP}_{sign} \rangle \times \prod_{i}^{m}\langle \text{SWAP}_{amp} \rangle_{i} \;. 
\end{eqnarray} 
The first part is the sign-dependent 
part of the SWAP-operator, the second part is itself a product over all contributions to the amplitude of SWAP. 
We find for the sign-dependent part:  
\begin{eqnarray}\label{eq12_supp} 
\langle \text{SWAP}_{sign} \rangle   
&=& \frac{\sum_{I I^{\prime}} 
|A(I, I^{\prime}, I^{s}, I^{\prime,s})|  
e^{i\Phi(I, I^{\prime})}} 
{\sum_{I, I{\prime}} |A(I, I^{\prime}, I^{s}, I^{\prime,s})| 
} \nonumber \\ 
&& 
\end{eqnarray} 
where we defined the weight 
\begin{eqnarray}\label{eq12a_supp} 
W(I, I^{\prime}) = |A(I, I^{\prime}, I^{s}, I^{\prime,s})| &=&   
|a_{I} a_{I^{\prime}} a_{I^{s}} a_{I^{\prime,s}}| \nonumber \\ 
&& 
\end{eqnarray} 
and $\Phi(I, I^{\prime})$ is the phase of $R$. The RVB wave function is real, thus, we have 
$e^{i\Phi(I, I^{\prime})} = \pm 1$. 

The amplitude-dependent part itself is a product. This allows us to express the quantity 
$\langle \text{SWAP}_{amp} \rangle$ to be evaluated as the product of a series of ratios 
so that the evaluation of each ratio only suffers from a much smaller fluctuation. 
This can practically be done be introducing $r_{i} \in [0,1]$ as a series of powers 
satisfying $r_{i} < r_{i+1}$, $r_{1} = 0$ and $r_{m+1} = 1$. Defining the weight 
\begin{eqnarray}\label{eq13_supp} 
\tilde{W}_{i}(I, I^{\prime}) &=&   
|A(I, I^{\prime}, I^{s}, I^{\prime,s})|^{1-r_{i+1}} \times 
\Big(|a_{I}|^{2} |a_{I^{\prime}}|^{2}\Big)^{r_{i+1}}\;, \nonumber \\
&&
\end{eqnarray}
we have 
\begin{eqnarray}\label{eq14_supp} 
\langle \text{SWAP}_{amp} \rangle = \prod_{i}^{m} \frac{\sum_{I, I^{\prime}} \tilde{W}_{i}(I, I^{\prime}) 
\times 
|R|^{r_{i+1} - r_{i}}} 
{\sum_{I, I^{\prime}} \tilde{W}_{i}(I, I^{\prime})}\;. 
\end{eqnarray}
If $r_{i+1} - r_{i}$ are chosen to be sufficiently small, 
each term in the above product can be evaluated easily and will 
not suffer from a relatively large error bar. 
It is easy to see that the crucial feature of the re-weighting scheme is that now the weights contain the 
amplitudes of the wave functions {\it after} the SWAP-operation. Consequently, all measurements are now non-zero. 
\begin{figure}[t]
\centering
\includegraphics[width=8.75cm]{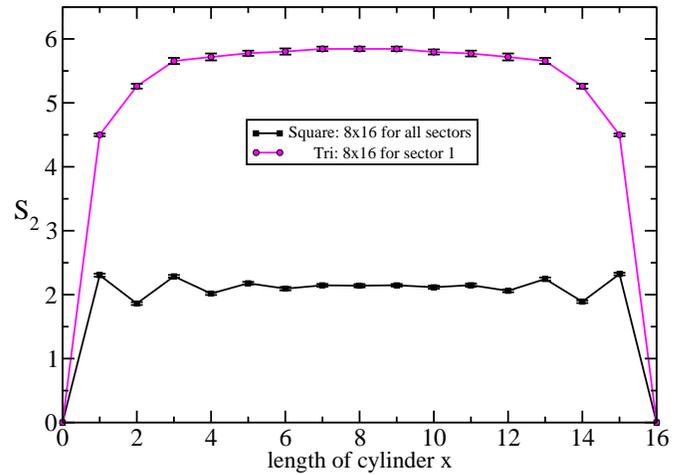}
\caption{ \label{triangular_square_supp}
SQUARE-TRIANGULAR: $S_{2}$ for the RVB wave function on the square and triangular (Tri) lattice is shown. 
The RVB wave function is sampled over all topological sectors for the square lattice, whereas in the triangular case, we fix the 
topological sector. 
The different behavior of the EE $S_{2}$ reflects the critical and gapped nature of the RVB state on the respective lattice. 
 }
\end{figure}

We point out that if we calculate the EE $S_{2}$ 
over a cylindrical area, 
using \eqref{eq11_supp} is sufficient. 
However, for a Levin-Wen construction the re-weighting scheme must be used in order to obtain sufficiently small error bars. 
For all Levin-Wen calculations in the Letter, a step width of $r_{i+1} - r_{i} = 0.2$ was used. 
Subsequently, the amplitude-dependent part of the SWAP-operator is a product \eqref{eq14_supp} consisting of $5$ factors. 
Adding a single calculation for the sign-dependent part of SWAP, we find that each Levin-Wen region $A$ requires six different calculations, 
the product of six partial expectation values corresponds to the ``full'' expectation value $\langle \text{SWAP} \rangle$. 

In the case of the RVB states on the triangular and kagome lattice, we remark that the error bars of all expectation values of the amplitude-part 
of the SWAP-operator converge relatively fast and are of order $10^{-4(5)}$. 
However, the error stemming from the sign-part $\langle \text{SWAP}_{sign}\rangle$ is significantly larger with values of $~0.2$ and $~0.3$ for 
the two lattices. 

As an example for an entanglement calculation without re-weighting scheme, we show the EE $S_{2}$ for a $12 \times 12$-square lattice 
for cylindrical regions in \Fig{compare_square_supp}. 
For bipartite lattices such as the square lattice, it is generally possible to sample over valence bond configurations, 
or perform a ``loop gas'' mapping \cite{sutherland}, 
without encountering a sign problem. Here, Marshall's sign rule provides a unique way of orientating all links, 
so that all statistical weights, which are given by valence bond state overlaps, are positive. 
For the bipartite square lattice, instead of four topological sectors, one has an extensive number of sectors. 
When using Valence Bond Monte Carlo (VBMC), it is possible to sample within a fixed sector, since here we start from a certain 
valence bond configuration and only perform local updates that can never lead into another sector. 

Contrarily, the Pfaffian MC cannot project onto a single sector on the square lattice, instead we sum over all topological sectors. 
However, we find that the behavior of the entanglement entropy $S_{2}$ does not depend on a specific sector 
or even the sum over all sectors (see \Fig{compare_square_supp}).

{\em 3) Comparison between the EE $S_{2}$ for the triangular and the square lattice and finite size effects.} -- 
\begin{figure}[t]
\centering
\includegraphics[width=8.75cm]{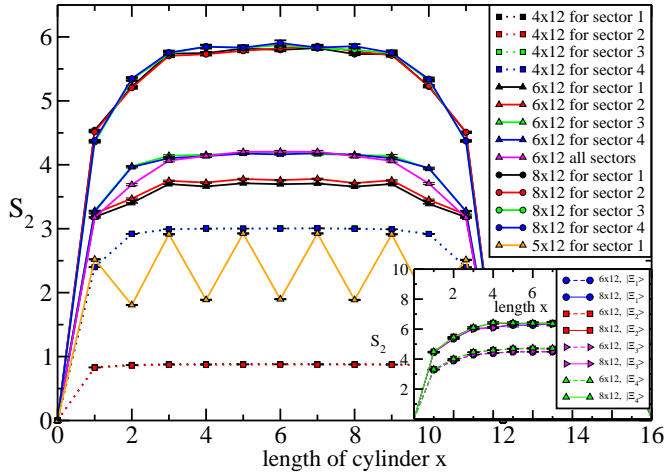}%{compare_468x12_1234_allsec_updated2-1}
\caption{ \label{kagome_sec14_supp}
TRIANGULAR: A comparison for $S_{2}$ for cylindrical regions between sectors $1, 2, 3$ and $4$ is shown. 
We observe that for large enough system size, the EE $S_{2}$ is the same for all four sectors. 
The inset shows $S_{2}$ for cylindrical regions for all four MES-states $|\Xi_{i}\rangle$, $i = 1, \ldots, 4$. 
It is evident that the four states $|\Xi_{i}\rangle$ become indistinguishable by their EE $S_{2}$ within errors for sufficiently large system sizes. 
 }
\end{figure}
We do a comparison of the EE $S_{2}$ for cylindrical regions $A$, 
via Pfaffian Monte Carlo on square and triangular lattices of the same size ($M \times 2N = 8 \times 16$). 
As discussed in the previous Section $2)$, the Pfaffian MC cannot project onto a single topological sector in the case of the square lattice. 
Hence, we sum over all sectors for the square lattice, and fix the topological sector for the triangular case. 
\Fig{triangular_square_supp} shows fundamentally different behavior in the EE $S_{2}$ for these two lattices. 
For the square lattice, we observe an ``even/odd'' or zig-zag effect  
that was investigated previously in Refs.\cite{ju} and \cite{stephan}. 
\Fig{triangular_square_supp} shows a weak oscillating behavior for growing cylinder length, 
which becomes much more pronounced in larger system sizes, and lattices that have a $M/2N$-ratio higher than $1/2$ \cite{ju,stephan}.  
The specific shape dependence of the even and odd branches is a reflection of the critical theory related to the 
square-lattice RVB wave function, which is conjectured to be related to a $2D$ quantum Lifshitz critical point \cite{stephan}. 

In contrast, $S_{2}$ obtained from the triangular lattice does not display critical behavior, including the absence of the 
even/odd effect.
\begin{figure}[t]
\centering
\includegraphics[width=8.75cm]{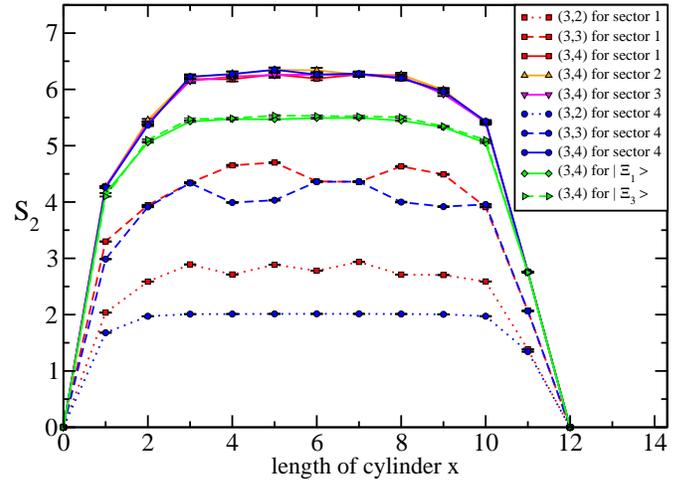}%{reaMxrealN_3x234_sec14-6}
\caption{ \label{kagome_sec4_supp2}
KAGOME: A comparison of $S_{2}$ for cylindrical regions between sectors $1, 2, 3$ and $4$ on the kagome lattice.
We observe that for large enough system sizes $(M,N) = (3,4)$ amounting to $72$ lattice sites,
the EE $S_{2}$ is the same for all four sectors within error bars.
Further, we present the EE $S_{2}$ for the MES-states by showing two examples corresponding to $|\Xi_{1}\rangle$ and $|\Xi_{3}\rangle$ for
$(M,N) = (3,4)$. The other two MES-states not shown here, $|\Xi_{2,4}\rangle$, exhibit the analogous behavior.
Since $S_{2}$ is the same for all topological sectors, we expect and confirm that $S_{2}$ is the same for all MES-states within errors
according to the MES-states ansatz in Eq.(3) in the Letter. }
\end{figure} 
Rather, $S_{2}$ saturates quickly away from cylinder lengths $x = 0$ and $x = 2N$, becoming independent of the system 
size $A$ (to within error bars). 
This ``flat'' behavior in the EE $S_{2}$ reflects the fact that the triangular lattice RVB wave function is 
the ground state of a gapped local Hamiltonian, in contrast with the square lattice case which is critical. 

\Fig{kagome_sec14_supp} shows more data on the triangular lattice. Shown is $S_{2}$ obtained for all four wave functions $|\Psi_{i}\rangle$, 
$i = 1, \ldots, 4$, 
and for the equal amplitude superposition of all four wave functions, $|\Psi_{\text{all sec}}\rangle = \sum_{i=1}^{4}|\Psi_{i}\rangle$, 
for lattices of sizes $M \times 2N = 4,5,6,8 \times 12$. 

%\begin{figure}[t]
%\centering
%\includegraphics[width=8.75cm]{reaMxrealN_3x234_sec14-7}
%\caption{ \label{kagome_sec4_supp2}
%KAGOME: A comparison of $S_{2}$ for cylindrical regions between sectors $1, 2, 3$ and $4$ on the kagome lattice. 
%We observe that for large enough system sizes $(M,N) = (3,4)$ amounting to $72$ lattice sites, 
%the EE $S_{2}$ is the same for all four sectors within error bars. 
%Further, we present the EE $S_{2}$ for the MES-states by showing two examples corresponding to $|\Xi_{1}\rangle$ and $|\Xi_{3}\rangle$ for 
%$(M,N) = (3,4)$. The other two MES-states not shown here, $|\Xi_{2,4}\rangle$, exhibit the analogous behavior. 
%Since $S_{2}$ is the same for all topological sectors, we expect and confirm that $S_{2}$ is the same for all MES-states within errors 
%according to the MES-states ansatz in Eq.(3) in the Letter. }
%\end{figure}
We note that we see a certain ``even/odd'' effect in $x$ for odd $M = 5$. This can be shown to be a finite size effect present for 
odd $M$ when $M \ll 2N$. It stems from the fact that for odd cylinder length $x$, there are less dimerizations corresponding to the   
Ising state that the sysem is in than there are for even length $x$. 
Thus, the values of $S_{2}$ at odd $x$ are larger than the ones at even $x$. 
The ``up/down''-pattern becomes less pronounced when we increase odd $M$ and finally vanishes. 
We stress that still the $S_{2}$ for odd (even) $x$ saturates and does not depend on odd (even) $x$ once a minimal size of region $A$ is reached. 

Furthermore, we notice differences in the respective $S_{2}$ belonging to a certain sector for sizes $4,6 \times 12$. 
Again, this is a finite size effect, and for a large enough system, e.g. $8 \times 12$, we do not observe differences in $S_{2}$ 
within error bars for different sectors. 

We also note that the equal amplitude superposition of all four sectors, $|\Psi_{\text{all sec}}\rangle$, 
should deliver the same EE $S_{2}$ as a single sector. 
The reason for that is that all states $|\Psi_{i}\rangle$, $i = 1, \ldots, 4$, and $|\Psi_{\text{all sec}}\rangle$ have the same 
TEE $-\gamma^{\prime}$ (see Section 4$)$). 

In \Fig{kagome_sec14_supp}, we observe that for a system of size $6 \times 12$, sectors $3$ and $4$ have the same EE $S_{2}$ within error bars 
as $|\Psi_{\text{all sec}}\rangle$. 
We conclude that in this case only the other two sectors, $1$ and $2$, suffer from finite size effects.  

The same effects can be found in the kagome lattice in Fig. $3$ in the Letter and in \Fig{kagome_sec4_supp2}. 
\Fig{kagome_sec4_supp2} shows that for the kagome lattice, we need a system size of $(M,N) = (3,4)$, amounting to $72$ sites, 
in order to achieve equality within error bars for the EE $S_{2}$ for all four sectors. 
Further, \Fig{kagome_sec4_supp2} shows that for $N \leq 4$ with fixed $M = 3$, we still see differences between the single sectors, 
including a zig-zag effect similar to the triangular lattice. 
Again, as in the triangular case, we see that $S_{2}$ also quickly experiences saturation once a minimum cylinder length $x$ is reached 
which strongly indicates that the Hamiltonian derived in Ref.\cite{seidel09}, which has the RVB state as its unique groundstate modulo 
topological degeneracy, is gapped.

{\em 4) Ground-state dependence of the TEE $-\gamma$.} -- 
The TEE $-\gamma$ is independent of the state used to calculate the Renyi entropy $S_{n}$ as long as the region $A$ is contractible. 
In the case that the area $A$ has at least one non-contractible boundary 
such as a cylindrical cut, $\gamma$ becomes state-dependent. 
If we now express any state in the basis of the so-called minimum entropy states (MES-states) \cite{zhang_ashvin}, 
$| \Psi_{\alpha} \rangle = \sum_{j} c_{j} |\Xi_{j} \rangle$, 
and take the area $A$ to be a torus with two non-contractible boundaries, then we have 
\begin{eqnarray}\label{gamma_torus_supp}
\gamma_{n}^{\prime}(\{p_{j}\}) = 2\gamma 
+ \frac{1}{n - 1}\text{ln}\bigg(\sum_{j} p_{j}^{n} d_{j}^{2(1-n)} \bigg) \;.
\end{eqnarray}
In the expression above, the sum over $j$ runs over all quasiparticles of the theory. 
The states $|\Xi_{j} \rangle$, $j = 1, \ldots, 4$, can be obtained by inserting a quasiparticle into the system. 
The $d_{j} \geq 1$ are the quantum dimensions of each quasiparticle and the $p_{j}$ in \eqref{gamma_torus_supp} 
are $p_{j} = |c_{j}|^2$. 
 
As stated previously, in this Letter we focus on the second Renyi entropy and set $n = 2$. 
Consequently, \eqref{gamma_torus_supp} simplifies to 
\begin{eqnarray}\label{gamma_torus2_supp}
\gamma^{\prime}(\{p_{j}\}) = 2\gamma + \text{ln}\bigg(\sum_{j} \frac{p_{j}^{2}}{d_{j}^{2}} \bigg) \;.
\end{eqnarray}
The numerical data shows the entanglement entropy to be independent of the wave functions $|\Psi_{i} \rangle$ used to calculate it 
for large enough systems for the kagome and the triangular lattice. 
Consequently, $\gamma^{\prime}$ has to be the same for all four states $|\Psi_{i} \rangle$, $i = 1, \ldots, 4$. 
We determined the value of $\gamma^{\prime}$ for both the triangular and the kagome to be close to $\text{ln}(2)$ if a single topological 
sector $|\Psi_{i} \rangle$, $i = 1, \ldots, 4$, is used to calculate it. 
In the Letter, we argued that the underlying quantum field theory must be an Abelian theory. Thus, all quasiparticles have the same quantum dimension $d = 1$. 
This implies that each wave function $|\Psi_{i} \rangle$ is a superposition of two MES-states: 
\begin{eqnarray}\label{mes_back0_supp}
|\Psi_{i} \rangle = \frac{1}{\sqrt{2}} \big(|\Xi_{a} \rangle \pm |\Xi_{b} \rangle \big) \;,
\end{eqnarray}
with $(a, b) = (1,2), (3,4)$. 
For later purposes, we also give $\gamma^{\prime}$ belonging to a single MES-state $|\Xi_{i} \rangle$, $i = 1, \ldots, 4$: 
\begin{eqnarray}\label{gammas0_supp}
|\Psi_{i} \rangle:\;\;\;\; \gamma^{\prime} &=& 2\;\text{ln}(2) + \text{ln}\bigg(\frac{1}{4} + \frac{1}{4}\bigg) = \text{ln}(2) \approx 0.69 \nonumber \\
|\Xi_{i} \rangle:\;\;\;\; \gamma^{\prime} &=& 2\;\text{ln}(2) + \text{ln}\big(1\big) = 2\;\text{ln}(2) \approx 1.38 \;. \nonumber \\
&&
\end{eqnarray}

{\em 5) MES-states for a dimer model on the triangular/kagome lattice.} -- 
The short-ranged dimer model on the kagome and the triangular lattice is known to be a $\mathbb{Z}_{2}$ liquid with toric code anyonic excitations. 
We will now derive the MES-states for this system. 
Note that we closely follow a similar derivation for the toric code model \cite{zhang_ashvin}. 

We recall that the $\mathbb{Z}_{2}$ dimer model has four topological sectors that are distinguished by the number of dimer cuts 
of a loop along each of the two directions of the lattice. 
For convenience, we temporarily use the subscripts '$0$' for an even of dimer cuts and '$1$' for an odd number, 
instead of '$e$' and '$o$', respectively. 
We only distinguish between even ($0$) and odd ($1$) number of cuts, thus, the four sectors are labeled: 
$(ab) = (00), (01), (10)$, and $(11)$. 
We denote a single sector by 
\begin{eqnarray}\label{sector_supp}
| \Psi_{ab} \rangle = \sum_{D \in (ab)}| D \rangle\;. 
\end{eqnarray}
Thus, in general, we can write a superposition of all four sectors as: 
\begin{eqnarray}\label{sector_all_supp}
| \Psi \rangle = \sum_{(ab)} c_{ab}| \Psi_{ab} \rangle \;. 
\end{eqnarray}
We proceed by doing a Schmidt decomposition of the states \eqref{sector_supp}. 
For this sake, we define a virtual cut $\Delta$ along the $\vec{a}$-direction (see Fig. $1$ in Letter) 
and define $|\Psi_{\{q_{l}\}b}^{A(B)} \rangle$ as the normalized equal superposition of all possible dimer coverings 
in subsystem $A(B)$. The boundaries defining the subsystems $A, B$ are along the $\vec{b}$-direction.  

The boundary condition connecting the two subsystems $A$ and $B$ is specified by $\{q_l = 0,1\}$, $l = 1, 2, \ldots, L$ 
($L$ = total length of boundaries) and the number of dimer crossings of the virtual cut $\Delta$ modulo $2$ 
equals $b = 0,1$. 
Consequently, the Schmidt-decomposed state is 
\begin{eqnarray}\label{schmidt_supp}
| \Psi_{ab} \rangle &=& \frac{1}{\sqrt{2N_q}}\sum_{\{q_l\} \in a}\bigg(| \Psi_{\{q_l\}0}^{A} \rangle | \Psi_{\{q_l\}b}^{B} \rangle \nonumber \\
&& \hspace{2.0cm} + | \Psi_{\{q_l\}1}^{A} \rangle | \Psi_{\{q_l\}b+1}^{B} \rangle\bigg)\;.\nonumber \\
&&
\end{eqnarray}
In the above, $\{q_l \in a = 0(1)\}$ denotes that only an even (odd) number of dimer crossings is allowed at the boundary $L_{1}$. 
The second boundary $L_{2}$ must have the same type, even or odd, of dimer cuts. 
The total length $L$ of the boundary is the combined length of $L_{1}$ and $L_{2}$. 
Subsequently, the total number of valid boundary conditions $\{q_l\} \in a$ in each parity sector is $N_q = 2^{L-2}$. 
We now replace all four terms in \eqref{sector_all_supp} with the corresponding Schmidt-decomposed state, form the outer product 
$| \Psi \rangle \langle \Psi| $ and execute the trace over the subspace $B$. 
This will give the reduced density matrix $\rho_{A}$: 
\begin{eqnarray}\label{ro_supp}
\rho_{A} &=& \frac{1}{\sqrt{2N_q}}\sum_{\{q_l \;even\}}
\bigg[(|c_{00}|^2 + |c_{01}|^2)\bigg( | \Psi_{\{q_l\}0}^{A} \rangle \langle \Psi_{\{q_l\}0}^{A} | \nonumber \\
&& \hspace{4.8cm} +\; | \Psi_{\{q_l\}1}^{A} \rangle \langle \Psi_{\{q_l\}1}^{A} |\bigg)\nonumber \\
&+&
2\text{Real}(c_{00}^{\ast}c_{01})\bigg( | \Psi_{\{q_l\}0}^{A} \rangle \langle \Psi_{\{q_l\}1}^{A} | + | \Psi_{\{q_l\}1}^{A} \rangle \langle \Psi_{\{q_l\}0}^{A} |\bigg) \bigg]\nonumber \\
&+&
\frac{1}{\sqrt{2N_q}}\sum_{\{q_l \;odd\}}
\bigg[(|c_{10}|^2 + |c_{11}|^2)\bigg( | \Psi_{\{q_l\}0}^{A} \rangle \langle \Psi_{\{q_l\}0}^{A} | \nonumber \\
&& \hspace{4.8cm} +\; | \Psi_{\{q_l\}1}^{A} \rangle \langle \Psi_{\{q_l\}1}^{A} |\bigg)\nonumber \\
&+&
2\text{Real}(c_{10}^{\ast}c_{11})\bigg( | \Psi_{\{q_l\}0}^{A} \rangle \langle \Psi_{\{q_l\}1}^{A} | + | \Psi_{\{q_l\}1}^{A} \rangle \langle \Psi_{\{q_l\}0}^{A} |\bigg) \bigg]\nonumber \\
&=& 
\frac{1}{2N_q}\sum_{\{q_l \;even\}}\bigg[ |c_{00}+c_{01}|^2 | \Psi_{\{q_l\}+}^{A} \rangle \langle \Psi_{\{q_l\}+}^{A} |\nonumber \\
&& \hspace{1.9cm} + |c_{00}-c_{01}|^2 | \Psi_{\{q_l\}-}^{A} \rangle \langle \Psi_{\{q_l\}-}^{A} |
                                   \bigg]\nonumber \\
&+& 
\frac{1}{2N_q}\sum_{\{q_l \;odd\}}\bigg[ |c_{10}+c_{11}|^2 | \Psi_{\{q_l\}+}^{A} \rangle \langle \Psi_{\{q_l\}+}^{A} |\nonumber \\
&& \hspace{1.9cm} + |c_{10}-c_{11}|^2 | \Psi_{\{q_l\}-}^{A} \rangle \langle \Psi_{\{q_l\}-}^{A} |
                                   \bigg]
\end{eqnarray}
with
\begin{eqnarray}\label{ro2_supp}
| \Psi_{\{q_l\}\pm}^{A} \rangle = \frac{1}{\sqrt{2}}(| \Psi_{\{q_l\}0}^{A} \rangle \pm | \Psi_{\{q_l\}1}^{A} \rangle). 
\end{eqnarray}
The two states \eqref{ro2_supp} are orthogonal to each other, as are the different dimer states that were summed over 
when the trace was taken in order to obtain $\rho_{A}$. 

Armed with this, we now calculate the $n$-th Renyi entropy $S_{n}$: 
\begin{eqnarray}\label{ee_supp}
S_{n} &=& \frac{1}{1-n}\text{ln}(\text{Tr}\rho_{A}^{n})\nonumber \\
&=& \frac{1}{n-1}\text{ln}\bigg[ \bigg(\frac{1}{2N_q}\bigg)^{2} N_{q} \bigg( (|c_{00}+c_{01}|^2)^n \nonumber \\
&& \hspace{3.2cm} + (|c_{00}-c_{01}|^2)^n \nonumber \\
&& \hspace{3.2cm} + (|c_{10}+c_{11}|^2)^n \nonumber \\
&& \hspace{3.2cm} + (|c_{10}-c_{11}|^2)^n \bigg)\bigg]\nonumber \\
&=& 
\text{ln}(N_q) + \frac{1}{1-n}\text{ln}\bigg(\sum_{j=1}^{4}p_{j}^{n}\bigg) \nonumber \\
&=& 
L\text{ln}(2) - \bigg[2\text{ln}(2) + \frac{1}{n-1}\text{ln}\bigg(\sum_{j}p_{j}^{n}\bigg)\bigg]
\end{eqnarray}
with
\begin{eqnarray}\label{4pj_supp}
&& p_{1} = \frac{|c_{00}+c_{01}|^{2}}{2} \;\;\;\;\;\;\;\;\; p_{2} = \frac{|c_{00}-c_{01}|^{2}}{2} \nonumber \\
&& p_{3} = \frac{|c_{10}+c_{11}|^{2}}{2} \;\;\;\;\;\;\;\;\; p_{4} = \frac{|c_{10}-c_{11}|^{2}}{2}\;.
\end{eqnarray}
Thus, up to an overall phase $e^{i\Phi_{j}}$, $j = 1, \ldots, 4$, the MES-states (that are orthogonal to each other) along $\vec{b}$ are each a 
superposition of two specific topological sectors: 
\begin{eqnarray}\label{MES_Z2_supp}
&&| \Xi_{1}\rangle = \frac{e^{i\Phi_{1}}}{\sqrt{2}} (| \Psi_{ee} \rangle + | \Psi_{eo} \rangle)\;,  \nonumber \\
&&| \Xi_{2}\rangle = \frac{e^{i\Phi_{2}}}{\sqrt{2}} (| \Psi_{ee} \rangle - | \Psi_{eo} \rangle)\;,  \nonumber \\
&&| \Xi_{3}\rangle = \frac{e^{i\Phi_{3}}}{\sqrt{2}} (| \Psi_{oo} \rangle + | \Psi_{oe} \rangle)\;,  \nonumber \\
&&| \Xi_{4}\rangle = \frac{e^{i\Phi_{4}}}{\sqrt{2}} (| \Psi_{oo} \rangle - | \Psi_{oe} \rangle)\;. 
\end{eqnarray}
We note that the analogous MES-states were previously found in Ref.\cite{zhang_ashvin} for the toric code model.  
In the case of the toric code, $|\Psi_{\{q_{l}\}b}^{A(B)} \rangle$ is defined to be the normalized equal superposition of all possible 
configurations of closed-loop strings $C$ in subsystem $A(B)$. 
Obviously, the degrees of freedom for the toric code model are the closed-loop strings $C$. 
For a dimer model, the above derivation of the MES-states is extremely similar to the toric code calculation once the degrees of freedom, 
the closed-loop strings, 
have been replaced by dimer degrees of freedom. 

There is an obvious correspondence between certain topological sectors of exactly solvable $\mathbb{Z}_{2}$ model systems, such as the 
toric code, the dimer model on the kagome and the triangular lattices, etc. $\ldots$, leading to a generic relation between these sectors 
and the MES-states, described by a matrix $T$ (see Section $6)$). 
To clarify this issue and especially to make clear that MES-states for double semion statistics have to be different, 
we will now investigate the transformation behavior of MES-states under modular transformations.

{\em 6) Modular $\mathcal{U}$- and $\mathcal{S}$-matrices and MES-states for the RVB spin liquid on the triangular/kagome lattice.} -- 
We will now determine the $\mathcal{U}$- and $\mathcal{S}$-matrices describing the quasiparticle statistics of the RVB system on the kagome 
and the triangular lattice by using the transformation behavior of the four (numerically confirmed) MES-states $|\Xi_{i} \rangle$, 
$i = 1, \ldots, 4$, 
under a certain modular transformation of the primitive lattice vectors $\vec{a}$ and $\vec{b}$. 

To start, we state that the $\mathcal{S}$- and $\mathcal{U}$-matrices describe the action of modular transformations on the degenerate 
ground-state manifold of the underlying topological quantum field theory on a torus. 
These two transformations generate the so-called modular group $SL(2, \mathbb{Z})$. 
All elements of this group can be generated by successive application of the following two generators $\mathfrak{s}$ and $\mathfrak{u}$, 
represented by the following matrix action on lattice vectors $\vec{a}, \vec{b}$: 
$S = \begin{pmatrix}
  \;\;\;0 &  1 \cr
 -1 &  0 \cr
\end{pmatrix}$ and 
$U = \begin{pmatrix}
  1 &  1 \cr
  0 &  1 \cr
\end{pmatrix}$.

Here, we refer the reader for more details to Ref.\cite{zhang_ashvin} 
and proceed by transforming 
the two primitive vectors $\vec{a}$ and $\vec{b}$ by applying $\mathfrak{us}^{-1}$ to them. 
This transformation, referred to as $R_{\pi/3}$ in the following, 
corresponds to: $\vec{a} \rightarrow \vec{a} - \vec{b}$ and $\vec{b} \rightarrow \vec{a}$. 
In this section, we consider kagome lattices of dimensions $(M,N) = (M,2M)$ and triangular lattices of size $(M,2N) = (2N,2N)$. 
For this choice of $(M,N)$ and $(M, 2N)$ for the respective lattice, we have as many links in $\vec{a}$-direction as in 
$\vec{b}$-direction and transforming the primitive vectors as described above corresponds to a rotation of the system by $\pi/3$. 
Hence, the name $R_{\pi/3}$.  

According to Ref.\cite{zhang_ashvin}, the overlaps $V_{ij} = \langle \Xi_{i}|R_{\pi/3}| \Xi_{j}\rangle$ 
between the bases $\{|\Xi_{i}\rangle\}$ and $\{R_{\pi/3}|\Xi_{j}\rangle\}$ form the unitary transformation 
$V = D^{\dagger} US^{-1} D$, where $D$ is a diagonal matrix of phases $D_{jj} = e^{i\Phi_{j}}$ corresponding to the 
phase freedom of choosing $|\Xi_{j}\rangle$. 

We now explicitly construct the $\mathcal{US}^{-1}$-matrices for toric code and for double semion quasiparticle statistics. 
The $\mathcal{U}$-matrix is a diagonal matrix with its $ii$th entry corresponding to the phase the $i$th quasiparticle acquires 
when it is exchanged with an identical one. 
Note that for the toric code anyon model, the quasiparticles consist of three bosons and one fermion. 
Thus, the self statistics yield phases of $1$ (bosons) and $-1$ (fermion). 
Contrarily, the double semion anyon model consists of two bosons, one semion and one anti-semion. 
In this case, the self statistics yield phases of $1$ (bosons), $i$ (semion) and $-i$ (anti-semion). 
 
For Abelian phases, the $ij$th entry of the $\mathcal{S}$-matrix corresponds to the phase the $i$th quasiparticle acquires when it 
encircles the $j$th quasiparticle. 

For the toric code, we have 
%\begin{eqnarray}
%\begin{split}
\begin{align}
\mathcal{U}=\begin{pmatrix}
 1     & 0     & 0     & \;\;\;0 \\
 0     & 1     & 0     & \;\;\;0 \\
 0     & 0     & 1     & \;\;\;0 \\
 0     & 0     & 0     & -1
\end{pmatrix},  \quad
\mathcal{S}=\frac{1}{2}\begin{pmatrix}
 1     & \;\;\;1     & \;\;\;1     & \;\;\;1 \\
 1     & \;\;\;1     & -1          & -1 \\
 1     & -1          & \;\;\;1     & -1 \\
 1     & -1          & -1          & \;\;\;1
\end{pmatrix}, 
\end{align}
% \end{split}
%\end{eqnarray}
while for the double semion phase, we have 
%\begin{eqnarray}
%\begin{split}
\begin{align}
\mathcal{U}=\begin{pmatrix}
 1     & 0     & 0     & \;\;\;0 \\
 0     & 1     & 0     & \;\;\;0 \\
 0     & 0     & i     & \;\;\;0 \\
 0     & 0     & 0     & -i
\end{pmatrix},  \quad
\mathcal{S}=\frac{1}{2}\begin{pmatrix}
 1     &  \;\;\;1     &  \;\;\;1     & \;\;\;1 \\
 1     &  \;\;\;1     & -1           & -1 \\
 1     & -1           & -1           & \;\;\;1 \\
 1     & -1           &  \;\;\;1     & -1
\end{pmatrix}. 
\end{align}
% \end{split}
%\end{eqnarray}

We now construct the $\mathcal{US}^{-1}$-matrices. In the case of toric code statistics, we have 
%\begin{eqnarray}
%\begin{split}
\begin{align}\label{USmatrix}
\mathcal{US}^{-1}=\frac{1}{2}\begin{pmatrix}
 \;\;\;1     & \;\;\;1     & \;\;\;1     & \;\;\;1 \\
 \;\;\;1     & \;\;\;1     & -1          & -1 \\
 \;\;\;1     & -1          & \;\;\;1     & -1 \\
 -1          & \;\;\;1     & \;\;\;1     & -1
\end{pmatrix}. 
% \end{split}
%\end{eqnarray}
\end{align}
In the case of double semion statistics, the $\mathcal{US}^{-1}$-matrix is given by
%\begin{eqnarray}
%\begin{split}
\begin{align}
\mathcal{US}^{-1}=\frac{1}{2}\begin{pmatrix}
 \;\;\;1     & \;\;\;1     & \;\;\;1     & \;\;\;1 \\
 \;\;\;1     & \;\;\;1     & -1          & -1 \\
 \;\;\;i     & -i          & -i          & \;\;\;i \\
 -i          & \;\;\;i     & -i          & \;\;\;i
\end{pmatrix}. 
% \end{split}
%\end{eqnarray}
\end{align}
The crucial difference lies therein  
that the $\mathcal{US}^{-1}$-matrix for the toric code quasiparticle statistics is a real matrix, 
contrarily the $\mathcal{US}^{-1}$-matrix for double semion statistics has real and complex entries. 
Thus, we conclude that $D$, the diagonal matrix of phases $D_{jj} = e^{i\Phi_{j}}$ corresponding to the 
phase freedom of choosing $|\Xi_{j}\rangle$, can be choosen to be real in the case of toric code statistics.  
This leads to a real matrix $V_{ij}$ in the basis of the MES-states. 

However, for double semion quasiparticle statistics, there exists no such choice of phases $D_{jj}$ that renders 
all overlap matrix elements $V_{ij}$ real in the MES-states basis. 
Thus, we showed that the MES-states \eqref{MES_Z2} are uniquely connected/bound to toric code statistics. 

To close this section, we derive the behavior of the four states $|\Psi_{\alpha \beta}\rangle$, $(\alpha \beta) = (ee), (eo), (oe), (oo)$ under 
$\pi/3$-rotation and then construct the matrix $T (T^{\dagger})$ which governs the change of basis from $|\Psi_{\alpha \beta}\rangle$-states to 
$|\Xi_{i}\rangle$-states (and vice versa).  
We recall that the lattices in this section are of sizes $(M,N) = (M,2M)$ for the kagome and $(M, 2N) = (2N,2N)$ for the triangular lattice,   
and now distinguish between an even 
and odd choice of $M$ for the kagome, and, respectively, between an even and odd choice of $N$ for the triangular lattice \cite{poilblanc2011competing}. 
We observe that if we choose $M(N)$ to be even for the kagome (triangular) lattice, the four topological sectors transform as 
\begin{eqnarray}\label{rot1_supp}
R_{\pi/3}| \Psi_{ee}\rangle = | \Psi_{ee}\rangle\,,\;\;  && R_{\pi/3}| \Psi_{eo}\rangle = | \Psi_{oe}\rangle \nonumber \\
R_{\pi/3}| \Psi_{oe}\rangle = | \Psi_{oo}\rangle\,,\;\;  && R_{\pi/3}| \Psi_{oo}\rangle = | \Psi_{eo}\rangle\;.
\end{eqnarray}
We now assume all phases $e^{i\Phi_{j}}$ in \eqref{MES_Z2} to be one. This will give the following basis transformation matrix $T$: 
\begin{align}
T=\frac{1}{\sqrt{2}}\begin{pmatrix}
 1     & \;\;\;1     & \;\;\;0     & \;\;\;0 \\
 1     & -1          & \;\;\;0     & \;\;\;0 \\
 0     & \;\;\;0     & \;\;\;1     & \;\;\;1 \\
 0     & \;\;\;0     & \;\;\;1     & -1
\end{pmatrix}. 
% \end{split}
%\end{eqnarray}
\end{align}
We note that $T^{\dagger} = T$ and proceed by constructing $T^{\dagger}R_{\pi/3}T$: 
\begin{eqnarray}
&& T^{\dagger}R_{\pi/3}T = \nonumber \\
&&\frac{1}{\sqrt{2}}\begin{pmatrix}
 1     & \;\;\;1     & 0     & \;\;\;0 \\
 1     & -1          & 0     & \;\;\;0 \\
 0     & \;\;\;0     & 1     & \;\;\;1 \\
 0     & \;\;\;0     & 1     & -1
\end{pmatrix}
\begin{pmatrix}
 1     & 0     & 0     & 0 \\
 0     & 0     & 1     & 0 \\
 0     & 0     & 0     & 1 \\
 0     & 1     & 0     & 0
\end{pmatrix}
\frac{1}{\sqrt{2}}\begin{pmatrix}
 1     & \;\;\;1     & 0     & \;\;\;0 \\
 1     & -1          & 0     & \;\;\;0 \\
 0     & \;\;\;0     & 1     & \;\;\;1 \\
 0     & \;\;\;0     & 1     & -1
\end{pmatrix}\nonumber \\
&&= 
\frac{1}{2}\begin{pmatrix}
 \;\;\;1     & \;\;\;1     & \;\;\;1     & \;\;\;1 \\
 \;\;\;1     & \;\;\;1     &      -1     &      -1 \\
 \;\;\;1     &      -1     & \;\;\;1     &      -1 \\
      -1     & \;\;\;1     & \;\;\;1     &      -1
\end{pmatrix}.
\label{find_US}
\end{eqnarray}
The matrix \eqref{find_US} is indeed the same as matrix \eqref{USmatrix}. 
Thus, our choice of phases $D_{jj} = e^{i\Phi_{j}} = 1$ for all four MES-states $|\Xi_{j}\rangle$, $j = 1, \ldots, 4$, listed in \eqref{MES_Z2} 
reproduces the $\mathcal{US}^{-1}$-matrix for toric code quasiparticle 
statistics \eqref{USmatrix}. 

Subsequently, we give the transformation behavior of the four topological sectors for odd $M(N)$ for the kagome (triangular) lattice: 
\begin{eqnarray}\label{rot2_supp}
R_{\pi/3}| \Psi_{ee}\rangle = | \Psi_{oe}\rangle\,,\;\;  && R_{\pi/3}| \Psi_{eo}\rangle = | \Psi_{ee}\rangle \nonumber \\
R_{\pi/3}| \Psi_{oe}\rangle = | \Psi_{eo}\rangle\,,\;\;  && R_{\pi/3}| \Psi_{oo}\rangle = | \Psi_{oo}\rangle\;.
\end{eqnarray}
One can now repeat the contruction of $T^{\dagger}R_{\pi/3}T$ in order to obtain the $\mathcal{US}^{-1}$-matrix in MES-basis again. 
In the case of odd $M(N)$ for the kagome (triangular) lattice, one can show that 
the appropriate phases, $D_{jj} = e^{i\Phi_{j}}$, are $D_{jj} = \{1,-1,1,-1\}$ for the MES-states $|\Xi_{j}\rangle$, $j = 1, \ldots, 4$, in \eqref{MES_Z2}.

%To summarize, we point out that the MES-states that we numerically confirmed to indeed apply to our triangular and to our kagome RVB systems rule 
%out the occurrence of double semion statistics. 
To summarize, these considerations show that our numerically confirmed MES-states for the kagome and triangular systems indeed imply 
the statistics of the $\mathbb{Z}$ phase, ruling out in particular double semion statistics. 
Thus, we have unambiguously identified the underlying quasiparticle statistics to be toric code statistics, 
taking into account that we already identified the topological entanglement entropy (TEE) $-\gamma$ 
to be $-\gamma = -\text{ln}(2)$ as expected for a topological $\mathbb{Z}_{2}$ spin liquid. 

After identifying the MES-states for the RVB state on the kagome and the triangular lattice, we note that 
all wave functions $|\Psi_{\alpha \beta} \rangle$, $(\alpha \beta) = (ee),(eo),(oe),(oo)$, 
are superpositions of the form 
\begin{eqnarray}\label{mes_back_supp}
|\Psi_{\alpha \beta} \rangle = \frac{1}{\sqrt{2}} \big(|\Xi_{a} \rangle \pm |\Xi_{b} \rangle \big) \;, 
\end{eqnarray} 
with $(a, b) = (1,2), (3,4)$. 

The numerical data shows the entanglement entropy $S_{2}$ to be independent of the sector and to be independent of 
the MES-state $|\Xi_{i} \rangle$, $i = 1, \ldots, 4$, used to calculate it, respectively.  
Again, this implies that $\gamma^{\prime}$ is the same for all $|\Psi_{\alpha \beta} \rangle$, $(\alpha,\beta) = (ee),(eo),(oe),(oo)$, 
and for all $| \Xi_{i}\rangle$, $i = 1, \ldots, 4$, respectively. 
A direct consequence of the latter is that all four quasiparticles belonging to the four MES-states must have the same quantum dimension $d_{i}$. 
Since every phase, Abelian and non-Abelian, has (at least) one quasiparticle with quantum dimension $d = 1$, we conclude that all 
quasiparticles have $d_{i} = 1$, $i = 1, \ldots, 4$. 
While this does not identify the topological phase, it again numerically confirms that the phase has to be Abelian. 

For reasons of completeness, we now give the other set of MES-states $\{|\Xi_{i}^{\prime}\rangle\}$ obtained for cuts along $\vec{a}$. 
For cuts along the other direction, $\vec{a}$, we obtain the corresponding MES-states from the action of the 
modular $\mathcal{S}$-matrix on the MES-states for cuts along the $\vec{b}$-direction: 
\begin{eqnarray}\label{MES_Z2a_supp}
&&| \Xi_{1}\rangle \longrightarrow \frac{1}{2} (| \Xi_{1}\rangle + | \Xi_{2}\rangle + | \Xi_{3}\rangle + | \Xi_{4}\rangle)\;,  \nonumber \\
&&| \Xi_{2}\rangle \longrightarrow \frac{1}{2} (| \Xi_{1}\rangle + | \Xi_{2}\rangle - | \Xi_{3}\rangle - | \Xi_{4}\rangle)\;,  \nonumber \\
&&| \Xi_{3}\rangle \longrightarrow \frac{1}{2} (| \Xi_{1}\rangle - | \Xi_{2}\rangle + | \Xi_{3}\rangle - | \Xi_{4}\rangle)\;,  \nonumber \\
&&| \Xi_{4}\rangle \longrightarrow \frac{1}{2} (| \Xi_{1}\rangle - | \Xi_{2}\rangle - | \Xi_{3}\rangle + | \Xi_{4}\rangle)\;. 
%&&  
\end{eqnarray}
leading to 
\begin{eqnarray}\label{MES_Z2b_supp}
&&| \Xi_{1}^{\prime}\rangle = \frac{1}{\sqrt{2}} (| \Psi_{oo} \rangle + | \Psi_{ee} \rangle)\;,  \nonumber \\
&&| \Xi_{2}^{\prime}\rangle = \frac{1}{\sqrt{2}} (| \Psi_{oo} \rangle - | \Psi_{ee} \rangle)\;,  \nonumber \\
&&| \Xi_{3}^{\prime}\rangle = \frac{1}{\sqrt{2}} (| \Psi_{eo} \rangle + | \Psi_{oe} \rangle)\;,  \nonumber \\
&&| \Xi_{4}^{\prime}\rangle = \frac{1}{\sqrt{2}} (| \Psi_{eo} \rangle - | \Psi_{oe} \rangle)\;. 
\end{eqnarray}
Here, we fixed the phases in \eqref{MES_Z2_supp} as seen/explained in Ref.\cite{zhang_ashvin}. 

%To close this section, we point out that it can be shown that any linear combination of sectors, 
%\begin{eqnarray}\label{linear_combination_supp}
%a_{1}|\Psi_{ee}\rangle + a_{2}|\Psi_{eo}\rangle + a_{3}|\Psi_{oe}\rangle + a_{4}|\Psi_{oo}\rangle \nonumber
%\end{eqnarray}
%with 
%$|a_{1}|^2 + |a_{2}|^2 + |a_{3}|^2 + |a_{4}|^2 = 1$, is either a MES-state or a linear combination of two or more MES-states. 
%This is so, since both $\{|\Psi_{i} \rangle\}$ and $\{|\Xi_{i} \rangle\}$, , $i = 1, \ldots, 4$, are bases and we can 
%(unitarily) transform back and forth between them. 
In order to extract the TEE $-\gamma^{\prime}$ from cylindrical bipartitions of the respective lattice, 
we checked the behavior of the EE $S_{2}$ for the MES-states and the respective so-called nonMES-states. 
%nonMES-states are all possible (normalized) superpositions of single sectors that do not correspond to a single MES-state 
%in the basis of the MES-states, but to a superposition of them. 
A nonMES-states is any linear combination of more than one MES-state.  
Here, we restrict ourselves to 
nonMES-states, referred to as $|\Sigma_{i}\rangle$ in the following, which are certain superpositions of exactly two topological sectors 
that are not leading to a single MES-state %that are defined to be states that are 
%superpositions of two sectors that do not form a MES-state 
from \eqref{MES_Z2b_supp}, e.g. 
\begin{eqnarray}\label{nonMES_Z2b_supp}
&&| \Sigma_{1}^{\prime}\rangle = \frac{1}{\sqrt{2}} (| \Psi_{ee} \rangle + | \Psi_{\alpha} \rangle)\;,  \nonumber \\
&&| \Sigma_{2}^{\prime}\rangle = \frac{1}{\sqrt{2}} (| \Psi_{ee} \rangle - | \Psi_{\alpha} \rangle)\;,  \nonumber \\
&&| \Sigma_{3}^{\prime}\rangle = \frac{1}{\sqrt{2}} (| \Psi_{eo} \rangle + | \Psi_{\beta} \rangle)\;,  \nonumber \\
&&| \Sigma_{4}^{\prime}\rangle = \frac{1}{\sqrt{2}} (| \Psi_{eo} \rangle - | \Psi_{\beta} \rangle)\;. 
\end{eqnarray}
with $\alpha = (eo),(oe)$ and $\beta = (oo),(ee)$. One can easily show that they are equal weight superpositions of all four 
MES-states \eqref{MES_Z2b_supp}. 
The state-dependent TEE vanishes in this case: $\gamma^{\prime} = 0$. 

The corresponding plots are shown in Fig. $3$ in the Letter, $S_{2}$ is the same for the two nonMES-states and it is larger than $S_{2}$ 
for the respective MES-state. 
Since we previously pointed out that $\gamma^{\prime}$ reaches its maximum $2\gamma$ (and $S_{2}$ its minimum) when calculating it with a MES-state, 
we can conclude that the difference for sufficiently long cylinder lengths $x$ between the MES-state and the nonMES-states of type 
\eqref{nonMES_Z2b_supp} is also $2\gamma$. This is confirmed by Fig. $3$ in the Letter. 

\bibliographystyle{apsrev}                                                 %{apsrev}  is the format you choose
\bibliography{together_paper_supplemental_material_April19_RogerJulia}{} %is the name of the bib file

\begin{thebibliography}{44}
\bibitem[{NNo()}]{NNote1}
\bibinfo{note}{Since there is a one-to-one correspondence between dimer
  coverings and singlet coverings, the dimer partition function is equivalent
  to the singlet (RVB) partition function.}
\end{thebibliography}
\end{document}